\documentclass[a4paper,12pt]{article}

\makeatletter
	\@addtoreset{equation}{section}
	\@addtoreset{figure}{section}
	\@addtoreset{table}{section}
\makeatother

\usepackage{amsmath,amsthm,amssymb,mathrsfs,amsfonts,mathtools,color,physics,url,listings,graphicx,tikz,here,multirow,array,ulem,bm,cite,hyperref}

\makeatletter
\let\MYcaption\@makecaption
\makeatother

\usepackage{subcaption}
\makeatletter
\let\@makecaption\MYcaption
\makeatother

\usetikzlibrary{calc,decorations.markings,intersections,arrows.meta}
\tikzset{->-/.style={decoration={markings,mark=at position .3 with {\arrow[scale=1.35]{latex}}},postaction={decorate}}}
\tikzset{-->-/.style={decoration={markings,mark=at position .9 with {\arrow[scale=1.35]{latex}}},postaction={decorate}}}
\tikzset{-<-/.style={decoration={markings,mark=at position .5 with {\arrow{latex reversed}}},postaction={decorate}}}

\newcommand{\Blue}[1]{\textcolor{blue}{#1}}

\begin{document}

\begin{titlepage}
\begin{flushright}
TIT/HEP-684\\
NORDITA 2021-032 \\
April,  2021
\end{flushright}
\vspace{0.5cm}
\begin{center}
{\Large \bf 
    WKB periods for higher order ODE and TBA equations
}

\lineskip .75em
\vskip 2.5cm

{\large  Katsushi Ito$^{a,}$\footnote{ito@th.phys.titech.ac.jp}, Takayasu Kondo$^{a,}$\footnote{t.kondo@th.phys.titech.ac.jp}, Kohei Kuroda$^{a,}$\footnote{k.kuroda@th.phys.titech.ac.jp} and  Hongfei Shu$^{b,c,}$\footnote{hongfei.shu@su.se } 
}
\vskip 2.5em
 {\normalsize\it 
 $^{a}$Department of Physics, Tokyo Institute of Technology,
Tokyo, 152-8551, Japan\\
$^{b}$
Beijing Institute of Mathematical Sciences and Applications (BIMSA), Beijing, 101408, China\\
$^{c}$ Nordita, KTH Royal Institute of Technology and Stockholm University\\
Roslagstullsbacken 23, SE-106 91 Stockholm, Sweden
}

\vskip 3.0em
\end{center}
\begin{abstract}
We study the WKB periods for the $(r+1)$-th order ordinary differential equation (ODE) which is obtained by the conformal limit of the linear problem associated with the 
$A_r^{(1)}$ affine Toda field equation. 
We compute the quantum corrections by using the Picard-Fuchs operators.
The ODE/IM correspondence provides a relation between the Wronskians of the solutions and the Y-functions which satisfy the thermodynamic Bethe ansatz (TBA) equation related to the Lie algebra $A_r$. 
For the quadratic potential, we propose a formula to show the equivalence between the logarithm of the Y-function and the WKB period, which is confirmed by solving the TBA equation numerically.
\end{abstract}

\end{titlepage}

\baselineskip=0.7cm

\section{Introduction}
In the exact WKB analysis, the wave function for the Schr\"odinger equation is expanded as the asymptotic series in the Planck constant.
The exact WKB periods defined by its Borel resummation lead to the energy spectrum of the system by employing the exact quantization condition \cite{bpv79,AIHPA_1983__39_3_211_0}.
The exact WKB periods are determined by their discontinuity structure and asymptotic behavior\cite{AIHPA_1999__71_1_1_0}. Recently, the relation between the exact WKB analysis and the integral equations satisfied by the exact WKB periods has been noticed in \cite{Ito:2018eon}.
The integral equations take the form of the Thermodynamic Bethe ansatz (TBA) equations of certain integrable models, and provide the solution to Voros's Riemann-Hilbert problem to determine the exact WKB period.
The TBA equations also provide an efficient method to solve the spectral problems with generic polynomial potential in quantum mechanics \cite{Ito:2018eon,Ito:2020ueb,Emery:2020qqu}.

For the Schr\"odinger equation with the monomial potential, the relation to the integrable models has been observed in \cite{Dorey:1998pt,Bazhanov:1998wj, Dorey:1999uk}, which is known as the ODE/IM correspondence, see \cite{Dorey:2007zx,Kuniba:2010ir,Dorey:2019ngq} for reviews.
One can construct the Y-functions using the Wronskians of the subdominant solutions of the ODE, which satisfy the Y-system.
The TBA equations are derived from the Y-system and the asymptotic behavior of the Y-functions \cite{Zamolodchikov:1991et,Ravanini:1992fi}.
The TBA equations for the monomial potential can be also obtained from those for generic potential by using the wall-crossing formula from the minimal chamber to the maximal chamber \cite{Gaiotto:2009hg,Alday:2010vh}.
Moreover, the logarithm of the Y-function is regarded as the (Borel resummed) WKB period, which has been seen in the gluon scattering amplitudes/minimal surface correspondence \cite{Alday:2010vh,Alday:2009dv, Hatsuda:2010cc}.

The ODE/IM correspondence has been generalized to a class of higher order ODEs \cite{Dorey:1999pv,Suzuki:1999hu,Dorey:2000ma,Suzuki:2000gi,Dorey:2006an,Gaiotto:2014bza}, which are realized as the conformal limit of the linear problem associated with the affine Toda field equations \cite{Dorey:2012bx,Ito:2013aea,Adamopoulou:2014fca, Ito:2015nla,Locke,kar72958,Ito:2018wgj,Ito:2020htm} \footnote{See also \cite{Lukyanov:2010rn,Fioravanti:2020udo,Sun:2012xw,Masoero:2015lga,Masoero:2015rcz,Ekhammar:2020enr} for related works.}.
It is interesting to explore the relation between the exact WKB periods for higher order ODE and the Y-functions for general integrable models.
The higher order ODEs also appear as the quantum Seiberg-Witten (SW) curves which describe the low-energy effective action of ${\cal N}=2$ supersymmetric gauge theories in the Nekrasov-Shatashvili limit of the Omega background \cite{Nekrasov:2009rc,Mironov:2009uv}.
In particular, for the Argyres-Douglas (AD) theory which is obtained by the scaling limit of the gauge theories\cite{Argyres:1995jj,Argyres:1995xn,Eguchi:1996vu}, the quantum SW curve becomes the higher order ODE\cite{Ito:2017ypt,Grassi:2018spf,Ito:2018hwp,Ito:2019twh,Ito:2020lyu}.
The quantum SW curves for a class of AD theories are labeled by a pair of Lie algebras $(G,G')$ \cite{Cecotti:2010fi}, where the quantum SW curve for the $(A_1,A_r)$-type AD theory corresponds to the second order ODE with $A_r$-type superpotential which is a polynomial potential of order $r+1$.

In \cite{Ito:2017ypt}, using the ODE/IM correspondence, the WKB periods for $(A_{1},A_{2r})$-type AD theories with a monomial potential is identified as the Y-functions of the $A_{2r}$-type Y-system \footnote{See also \cite{Grassi:2019coc,Fioravanti:2019vxi,Fioravanti:2019awr} for the case of $SU(N)$ ${\cal N}=2$ gauge theory. }.
The effective central charge of the associated conformal field theory has been calculated from the TBA equations.
An interesting feature is that the conformal field theory has the same central charge predicted from the 2d/4d correspondence \cite{Beem:2013sza}.
Moreover, in the asymptotic expansions of the Y-functions, the coefficients are identified with the eigenvalues of the integrals of motion of the conformal field theory \cite{Bazhanov:1998wj,Bazhanov:2001xm}.

In this paper, we study the relation between the WKB periods and the Y-functions for the higher order ODE using the ODE/IM correspondence.
The exact WKB method for general higher order ODE has been developed in \cite{honda2015virtual,doi:10.1063/1.525467}, where the Stokes phenomena of the WKB solutions are extensively studied.
In \cite{Neitzke:2017yos,Hollands:2019wbr,Dumas:2020zoz,Yan:2020kkb}, using the spectral network \cite{Gaiotto:2012rg}, the relation between the WKB periods and the TBA equations is examined for the Hitchin system.
One can identify the cycle of the WKB period associated with the Y-function from the Stokes graph.

In this paper, we focus on a class of the ODE obtained by replacing the second order derivative in the Schr\"odinger equation with a higher order derivative.
For the $(N+1)$-th order potential, the $(r+1)$-th order ODE describes the quantum SW curve of the $(A_r,A_N)$-type AD theory.
For this type of ODE, one can compute quantum corrections to the classical period similar to the Schr\"odinger equation.
In particular, for a quadratic potential, one calculates the quantum corrections by acting the differential operators on a set of classical periods of the WKB curve.
From the ODE/IM correspondence, the corresponding Y-functions can be obtained from the cross ratios of the Wronskians of the subdominant solutions, which satisfy the $A_r$-type Y-system.
The TBA equation gives the asymptotic expansion of the Y-function in the Planck constant.
One can also compute the coefficients numerically in the series.
Comparison of the coefficients with the WKB periods provides a non-trivial test of the ODE/IM correspondence.
We will examine the correspondence for the higher order ODE with the quadratic potential. 
As a generalization, we will also study the correspondence for the cubic potential of the third order ODE, which shows the wall-crossing of the TBA equations.
%A generalization to higher order potential will be studied in a subsequent paper \cite{ItKoKuSh3}.

This paper is organized as follows.
In section \ref{sec:WKB period}, we apply the WKB analysis to the higher order ODE.
Using the Riccati equation, we calculate the quantum corrections to the classical WKB period.
In section \ref{sec:Y-function}, we focus on the $(r + 1)$-th order ODE with a monomial potential of order $N+1$.
We define the Y-functions and the Y-system from the Wronskian of the solutions.
We discuss the semi-classical limit of the Y-function, where the logarithm of the Y-function is represented by the WKB period associated with the cycle on the WKB curve.
In section \ref{sec:TBA}, we consider the TBA equations and the asymptotic expansion of the Y-functions in the large spectral parameter.
We confirm the quantum corrections to the WKB period agree with the coefficients of the expansion of the Y-function numerically.
We also discuss the PNP (perturbative-non-perturbative) relations for the WKB periods.
In section \ref{sec:wall_crossing}, the wall-crossing phenomena of the TBA equations for higher order ODE with polynomial potential are discussed.
Section \ref{sec:conclusions} is devoted to conclusions and discussion.

\section{WKB analysis for higher order ODE}
\label{sec:WKB period}
In this section, we apply the WKB analysis to the higher order ODE related to the conformal limit of the linear problem associated with the affine Toda field equation for the $A_r^{(1)}$-type affine Lie algebra \cite{Ito:2013aea,Adamopoulou:2014fca}.
It is also regarded as the quantum SW curve for the $(A_r,A_N)$-type Argyres-Douglas theory \cite{Ito:2017ypt}.
The ODE is the $(r+1)$-th order equation defined in the complex plane:
\begin{equation}
    \qty[(-1)^r \epsilon^{r+1} \partial_x^{r+1} +  p(x,\{u_i\})]\psi(x,\{u_i\};\epsilon)=0,\qquad p(x,\{u_i\})=\sum_{i=0}^{N+1} u_{N+1-i} x^i.
    \label{eq:higher_order_ODE}
\end{equation}
We call $p(x,\{u_i\})$ the potential term because in the Schr\"odinger case, $p(x)$ represents potential minus energy.
The coefficients $u_i$ and $\epsilon$ are complex parameters, where $\epsilon$ plays a role of the Planck constant.
The WKB solution takes the form
\begin{equation}
    \psi(x,\{u_i\};\epsilon) = \exp\qty(\frac{1}{\epsilon}\int^xS(x^{\prime},\{u_i\};\epsilon)\dd x^{\prime}),
    \label{eq:WKB_ansatz}
\end{equation}
where $S(x,\{u_i\};\epsilon)$ is expanded in $\epsilon$ as
\begin{equation}
    S(x,\{u_i\};\epsilon) = \sum_{n=0}^{\infty} \epsilon^n S_{n}(x,\{ u_i\}).
    \label{eq:formal_series_exp_S}
\end{equation}
The $S(x,\{u_i\};\epsilon)$ in \eqref{eq:WKB_ansatz} satisfies the higher order version of the Riccati equation:
\begin{equation}
    (-1)^r\qty(\epsilon\partial_x + S(x,\{u_i\};\epsilon))^rS(x,\{u_i\};\epsilon) + p(x,\{u_i\}) = 0.
    \label{eq:Riccati_eq_Ar}
\end{equation}
Substituting \eqref{eq:formal_series_exp_S} into \eqref{eq:Riccati_eq_Ar} and expanding it in $\epsilon$, one can determine $S_n$ recursively. 
For example, $O(\epsilon^0)$ term in \eqref{eq:Riccati_eq_Ar} is 
\begin{equation}
(-S_0)^{r+1} - p=0,
\end{equation}
from which we find
\begin{equation}
    S_0 = -e^{\frac{2\pi i k}{r+1}}p^{\frac{1}{r+1}},\qquad k=0,1,\dots, r.
\end{equation}
We set $k=0$ in the following calculation.
The higher order correction terms can be obtained recursively and expressed by $S_0$:
\begin{equation}
    S_1 = -\frac{r}{2}\frac{\partial_x S_0}{S_0},\qquad 
    S_2 = \frac{r(r+2)}{12}\qty(\frac{\partial_x^2 S_0}{S_0^2} - \frac{3}{2}\frac{\qty(\partial_x S_0)^2}{S_0^3}),\dots.
\end{equation}
Formulas for $S_4$, $S_6$ and $S_8$ are presented in Appendix \ref{sec:wkb}.
We have computed $S_n$ $(n\leq 20)$ for $r\leq 7$.
From our analysis, we observe that $S_n$ for odd $n$ become the total derivative as in the case of the Schr\"odinger equation.
For even $r$, we also find $S_n$ for $n=2(r+1)k+r+2$ $(k=0,1,\dots)$ become the total derivative, which is new for higher order ODE.
As we will see in section \ref{sec:derivation_of_TBA}, these vanishing terms can be seen also from the TBA equations.

We next study the WKB period for the ODE \eqref{eq:higher_order_ODE}. 
We introduce the WKB curve:
\begin{equation}
    \Sigma\;\colon\; y^{r+1}=p(x,\{u_i\}).
    \label{eq:WKB_curve}
\end{equation}
The curve \eqref{eq:WKB_curve} represents $(r+1)$-fold covering of the complex plane with $N+1$ branch points. On this curve, there is a basis of meromorphic differentials of the form:
\begin{equation}
    \frac{x^{i-1}}{y^a}dx, \quad i=1,\dots, N-1, \ a=1,\dots, r.
    \label{eq:merodiff1}
\end{equation}
We also introduce the set of the SW differentials $y^a dx$ $(a=1,\dots, r)$, which generate the basis \eqref{eq:merodiff1}:
\begin{equation}
    \partial_{u_i}(y^a dx)=\frac{a}{r+1}\frac{x^{N-i+1}}{ y^{r+1-a}}dx.
    \label{eq:sw1}
\end{equation}
We now define the WKB period $\Pi_{\gamma}(\{u_i\};\epsilon)$ by 
\begin{equation}
    \Pi_{\gamma}(\{u_i\};\epsilon) \coloneqq \int_{\gamma} S(x,\{u_i\};\epsilon)\dd x.
    \label{eq:qp1}
\end{equation}
where $\gamma$ is a $1$-cycle on the WKB curve \eqref{eq:WKB_curve}.
Substituting the formal expansion \eqref{eq:formal_series_exp_S} into \eqref{eq:qp1}, we obtain
\begin{equation}
    \Pi_{\gamma}(\{u_i\};\epsilon) = \sum_{n=0}^{\infty} \epsilon^n \Pi_{\gamma}^{(n)}(\{u_i\}),
\end{equation}
where $\Pi^{(n)}_{\gamma}(\{u_i\})$ is defined by
\begin{equation}
    \Pi^{(n)}_{\gamma}(\{u_i\})\coloneqq \int_{\gamma} S_{n}(x,\{u_i\})\dd x,\qquad n=0,1,2,\dots.
    \label{eq:def_q_correction}
\end{equation}
$\Pi^{(0)}_{\gamma}$ is called the classical WKB period and $\Pi^{(n)}_{\gamma}(n\geq1)$ represents the quantum correction to the classical WKB period.
Since the period integrals of total derivative term vanish, we can ignore the total derivative terms in $S_n$.
Then we find $\Pi^{(n)}_\gamma=0$ for odd $n$ and some even $n$.
For $S_n$ which is not written as the total derivative, we express $S_n(x,\{u_i\})$ by the linear combination of the basis \eqref{eq:merodiff1} up to total derivatives. 
Then by using \eqref{eq:sw1}, $S_n$ can be expressed in terms of the derivative of $y^{l_n}=p^{l_n/(r+1)}$ with respect to $u_i$ for some $1\leq l_n\leq r$:
\begin{equation}
    S_n(x,\{u_i\}) = \mathcal{O}^{\mathrm{PF}}_n(\{u_i\},\{\partial_{u_i}\})\; \qty(p(x,\{u_i\}))^{\frac{l_n}{r+1}}+d(*),\qquad l_n\in\{1,\dots,r\},
    \label{eq:def_PF_operator}
\end{equation}
where $d(*)$ represents the total derivatives.
We refer $\mathcal{O}^{\mathrm{PF}}_n(\{u_i\},\{\partial_{u_i}\})$ as the Picard-Fuchs operator.
For example, $S_2$ is expressed as
\begin{equation}
    S_2=-\frac{r+1}{24}
    \sum_{j=0}^{N-1} (N+1-j)(N-j)u_i \partial_{u_{N+1}}\partial_{u_{j+2}}y^{r}+d(*).
\end{equation}
Substituting \eqref{eq:def_PF_operator} into \eqref{eq:def_q_correction}, and changing the ordering of the integral with respect to $x$ and the differential $\mathcal{O}^{\mathrm{PF}}_n(\{u_i\},\{\partial_{u_i}\})$, one can compute the quantum corrections from the period integral of $p^{l_n/(r+1)}$:
\begin{equation}
    \Pi^{(n)}_{\gamma}(\{u_i\}) = \mathcal{O}^{\mathrm{PF}}_n(\{u_i\},\{\partial_{u_i}\}) \hat{\Pi}^{(0),l_n}_{\gamma}(\{u_i\}),
\end{equation}
where we defined $\hat{\Pi}^{(0),l_n}_{\gamma}$ by
\begin{equation}
    \hat{\Pi}^{(0),l_n}_{\gamma}(\{u_i\}) = \oint_{\gamma} \qty(p(x,\{u_i\}))^{\frac{l_n}{r+1}}\dd x.
\end{equation}
Note that $-\hat{\Pi}^{(0),1}_{\gamma}(\{u_i\})$ is the classical WKB period $\Pi^{(0)}_{\gamma}(\{u_i\})$ and $\hat{\Pi}^{(0),l_n}_{\gamma}$ is the period integral of the SW differential $y^{l_n}dx$ on the cycle. We refer it as the classical SW period.

\subsection{Quadratic potential}
\begin{figure}[H]
    \centering
    \footnotesize
    \begin{tikzpicture}[]
        \def\ra{2.5};
        \def\dv{0.07};
        \def\rt{1.5};
        \def\rv{0.8};
        \coordinate[] (ni) at (-4, 0);
        \coordinate[] (pi) at (4, 0);
        \coordinate[label=above:$-\sqrt{u}$] (nx) at (-1.5, 0);
        \coordinate[label=above:$\sqrt{u}$] (px) at (1.5, 0);
        \coordinate[] (g1np) at ({-\ra}, {\dv});
        \coordinate[] (g1nn) at ({-\ra}, {-\dv});
        \coordinate[] (g1pp) at ({\ra}, {\dv});
        \coordinate[] (g1pn) at ({\ra}, {-\dv});
        
        \coordinate[] (g1inp) at ({-\rt}, {\rv});
        \coordinate[] (g1ipn) at ({\rt}, {-\rv});
        \coordinate[] (g1ipp) at ({\rt}, {\rv});
        \coordinate[] (g1inn) at ({-\rt}, {-\rv});
    
        \fill (nx) circle [radius=2pt, red];
        \fill (px) circle [radius=2pt, red];
        
        \draw[line width=0.4mm, gray, dashed] (ni) -- (nx);
        \draw[line width=0.4mm, gray, dashed] (px) -- (pi);
        
        \draw[line width=0.3mm, red] (g1np) to [out=85, in=180] (g1inp);
        \draw[line width=0.3mm, red, ->-] (g1inp) to [out=0, in=180] (g1ipn);
        \draw[line width=0.3mm, red] (g1ipn) to [out=0, in=265] (g1pn);
        
        \draw[line width=0.3mm, dashed, red] (g1pp) to [out=95, in=0] (g1ipp);
        \draw[line width=0.3mm, dashed, red, ->-] (g1ipp) to [out=180, in=0] (g1inn);
        \draw[line width=0.3mm, red, dashed] (g1inn) to [out=180, in=275] (g1nn);
    \end{tikzpicture}
    \caption{The cycle $\gamma_m$. The solid line is on the $m$-th sheet while the dashed line is on the $(m+1)$-th sheet of the WKB curve $\Sigma$.}
    \label{fig:cycle_gamma_m}
\end{figure}
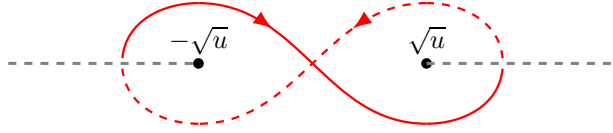

Computation of the classical periods for a general polynomial $p(x)$ is a highly non-trivial problem.
In this paper, we consider the case where $p(x)$ is quadratic in $x$.
By the shift of $x$, we can set $p(x)$ as 
\begin{equation}
    p(x,u_0,u_2) = u_0 x^2 + u_2.
    \label{eq:quadratic_potential}
\end{equation}
To calculate the quantum corrections, we first compute the classical SW period $\hat{\Pi}_{\gamma}^{(0),l}$.
The WKB curve $\Sigma$ is a $(r+1)$-sheeted cover of $\mathbb{C}$ with two branch points $x=\pm\sqrt{u}$ $(u=-u_2/u_0)$. 
We set one of the branch cuts to be the half-line which ends on $x=\sqrt{u}$, while the other cut ends on $x=-\sqrt{u}$.
Let us define the sheet labeled by $m\in\{1,\dots,r+1\}$, where $y$ is given by
\begin{equation}
    y_m = e^{\frac{2\pi i m}{r+1}} \qty( u_0 x^2 + u_2)^{\frac{1}{r+1}},\qquad m=1,\cdots,r+1.
\end{equation}
We also introduce the basis $\gamma_m\;(m=1,\dots,r+1)$ of the $1$-cycles on $\Sigma$, which encircle one of the branch point clockwise and the other anti-clockwise respectively over the $m$-th and the $(m+1)$-th sheets as in Figure \ref{fig:cycle_gamma_m}.
The intersection number of $\gamma_m$ to $\gamma_{m\pm1}$ is $\pm1$.
\noindent
For the cycle $\gamma_m$, $\hat{\Pi}_{\gamma_m}^{(0),l_n}$ becomes
\begin{equation}
    \hat{\Pi}^{(0),l_n}_{\gamma_m}
    =\qty(e^{\frac{2\pi iml_n}{r+1}}-e^{\frac{2\pi i(m+1)l_n}{r+1}})\int_{-u^{1/2}}^{u^{1/2}}\qty(u_0 x^2 +u_2)^{\frac{l_n}{r+1}}\dd x.
\end{equation}
By change of the variable, we obtain
\begin{equation}
    \hat{\Pi}^{(0),l_n}_{\gamma_m} = 2 e^{\frac{\pi il_n}{r+1}(2m+3)}\sin\qty(\frac{\pi l_n}{r+1}) u_2^{\frac{1}{2}+\frac{l_n}{r+1}} u_0^{-\frac{1}{2}} B\qty(\frac{1}{2},1+\frac{l_n}{r+1}),
\end{equation}
where $B(a,b)$ is the beta function defined by
\begin{equation}
    B(x,y)\coloneqq \int_0^1 t^{x-1}(1-t)^{y-1}\dd t = \frac{\Gamma\qty(x)\Gamma\qty(y)}{\Gamma\qty(x+y)},\qquad \Re{x}>0,\Re{y}>0.
\end{equation}

\subsection{An example: the third order ODE}
\label{sec:third_order_case}
We now perform the WKB analysis for the third order ODE as an example.
The Riccati equation \eqref{eq:Riccati_eq_Ar} for $r=2$ reads
\begin{equation}
    \epsilon^2 \partial_x^2S(x,\{u_i\};\epsilon)+3 \epsilon  S(x,\{u_i\};\epsilon) \partial_xS(x,\{u_i\};\epsilon)+S(x,\{u_i\};\epsilon)^3 +  p(x,\{u_i\}) = 0.
\end{equation}
Substituting the expansion \eqref{eq:formal_series_exp_S}, we can solve $S_n$ in terms of $S_0$. 
We see that $S_{n}$ for odd $n$ and $S_{6k+4}$ $(k=0,1,\dots)$ become total derivatives by explicit calculation\footnote{We confirmed this up to $n\leq 20$}.
Here we show the list of $S_{n}$ for even $n$ up to $n=10$:
\begin{equation}
\begin{aligned}
    &S_{0}=-p^{1/3},\\
    &S_2=-\frac{1}{36}\frac{\partial_x^2p}{p^{4/3}}+d(*),\\
    &S_4=d(*),\\
    &S_6=\frac{979 \qty(\partial_x^2p)^3}{122472 p^{14/3}}-\frac{55 \partial_x^4p\partial_x^2 p}{13608 p^{11/3}}-\frac{4 \qty(\partial_x^3p)^{2}}{5103 p^{11/3}}+\frac{5 \partial_x^6p}{15552 p^{8/3}}+d(*),\\
    &S_8=-\frac{2743 \qty(\partial_x^2p)^4}{157464 p^{19/3}}+\frac{1625 \partial_x^4p\qty(\partial_x^2 p)^2}{139968 p^{16/3}}-\frac{65 \partial_x^6p \partial_x^2p}{69984 p(x)^{13/3}}\\
    &\qquad\qquad +\frac{143 \qty(\partial_x^3p)^2 \partial_x^2p}{52488 p^{16/3}}-\frac{83 \qty(\partial_x^4p)^2}{93312 p^{13/3}}-\frac{5 \partial_x^3p \partial_x^5p}{8748 p^{13/3}}-\frac{7 \partial_x^8p}{186624 p^{10/3}}+d(*),\\
    &S_{10}=d(*).
\end{aligned}
\label{eq:A2_Sn}
\end{equation}
We then calculate the quantum corrections $\Pi^{(n)}_{\gamma}$ which are non-zero only for $n=0,2\mod6$. 
The integer $l_n$ in \eqref{eq:def_PF_operator} is given by
\begin{equation}
    l_n=\begin{cases}
        1 & \textrm{for} \quad n = 0 \mod 6,\\
        2 & \textrm{for} \quad n = 2 \mod 6.
        \end{cases}
\end{equation}
The Picard-Fuchs operators  $\mathcal{O}^{\mathrm{PF}}_n(u_0,u_2,\partial_{u_2})$ up to $\epsilon^{18}$ take the following form:
\begin{equation}
\begin{aligned}
    \mathcal{O}^{\mathrm{PF}}_2&=\frac{u_0}{4}\partial_{u_2}^2, & 
    \mathcal{O}^{\mathrm{PF}}_6&=\frac{89 u_0^3}{5040}\partial_{u_2}^5, &
    \mathcal{O}^{\mathrm{PF}}_8&=-\frac{211 u_0^4}{40320}\partial_{u_2}^7, \\
    \mathcal{O}^{\mathrm{PF}}_{12}&=-\frac{2160997 u_0^6}{3632428800}\partial_{u_2}^{10}, &
    \mathcal{O}^{\mathrm{PF}}_{14}&=\frac{897629 u_0^7}{4151347200}\partial_{u_2}^{12}, &
    \mathcal{O}^{\mathrm{PF}}_{18}&=\frac{26543159161 u_0^9}{779776284672000}\partial_{u_2}^{15}. \\
\end{aligned}
\end{equation}
Then, the ratios of the quantum corrections to the classical SW periods are
\begin{equation}
\begin{aligned}
    \frac{\Pi^{(2)}_{\gamma}}{\hat{\Pi}^{(0),2}_{\gamma}}&=\frac{7 u_0}{144 u_2^2}, &   
    \frac{\Pi^{(6)}_{\gamma}}{\hat{\Pi}^{(0),1}_{\gamma}}&=\frac{21983 u_0^3}{1119744 u_2^5},\\
    \frac{\Pi^{(8)}_{\gamma}}{\hat{\Pi}^{(0),2}_{\gamma}}&=\frac{26317819 u_0^4}{322486272 u_2^7}, &
    \frac{\Pi^{(12)}_{\gamma}}{\hat{\Pi}^{(0),1}_{\gamma}}&=\frac{70877384469605 u_0^6}{13792092880896 u_2^{10}}, \\
    \frac{\Pi^{(14)}_{\gamma}}{\hat{\Pi}^{(0),2}_{\gamma}}&=\frac{429318166799748793 u_0^7}{4694326886006784 u_2^{12}}, &   
    \frac{\Pi^{(18)}_{\gamma}}{\hat{\Pi}^{(0),1}_{\gamma}}&=\frac{14039462154947603772784295 u_0^9}{286408827560773287936 u_2^{15}}.\\
\end{aligned}
\end{equation}
In appendix \ref{sec:PF_op_q_corrections}, we will show the Picard-Fuchs operators and the quantum corrections for the case from the fourth to the seventh order ODE.

\section{Y-functions and WKB periods}
\label{sec:Y-function}
In this section, we first review the ODE/IM correspondence to define the Y-functions and the Y-system from the higher order ODE \cite{Dorey:2007zx}. We then discuss the relation between the Y-functions and the WKB periods for the higher order ODE, which has been known for the second order ODE in \cite{Ito:2017ypt}.

\subsection{ODE/IM correspondence}\label{sec:T_Y_system}
Let us consider the $(r+1)$-th order ODE with the monomial potential:
\begin{equation}
    \qty[(-1)^{r}\epsilon^{r+1}\partial_x^{r+1} +  
    \qty( u_{0} x^{N+1} + u_{N+1})]\psi(x,u_0,u_{N+1};\epsilon) = 0.
    \label{eq:Ar_ODE}
\end{equation}
The ODE/IM correspondence for (\ref{eq:Ar_ODE}) has been studied in \cite{Dorey:2000ma,Suzuki:1999hu, Ito:2017ypt}.
This ODE is invariant under the Symanzik (Sibuya) \cite{sibuya} rotation $(x,u_0,u_{N+1};\epsilon)\to (\omega^{-1} x,u_{0},\omega^{-(N+1)} u_{N+1};\epsilon)$ with $\omega=e^{\frac{2\pi i}{N+h+1}}$ and $h=r+1$. 
For our purpose, it is more convenient to regard the Symanzik rotation as the transformation of $\epsilon$, i.e. $(x,u_0,u_{N+1};\epsilon)\to (x,u_0,u_{N+1};e^{\frac{2\pi i}{h}}\epsilon)$.
Since the rotated solution satisfies the same ODE, this rotational symmetry enables us to generate solutions from a given one.

Let us consider a solution $\phi(x,u_0,u_N;\epsilon)$ to the ODE \eqref{eq:Ar_ODE} whose asymptotic behavior is given by
\begin{equation}
    \phi(x,u_{0},u_{N};\epsilon)\sim\frac{\epsilon^{\frac{r}{2}}u_{0}^{-\frac{r}{2h}}}{i^{\frac{r}{2}}\sqrt{h}}x^{-\frac{r(N+1)}{2h}}\exp\qty(-\frac{1}{\epsilon}\frac{u_{0}^{\frac{1}{h}}h}{N+h+1}x^{\frac{N+h+1}{h}}),\quad \abs{x}\to\infty,
\end{equation}
along the positive real axis.
The normalization factor is fixed for later convenience.
Note that $\phi(x,u_0,u_{N+1};\epsilon)$ is the subdominant solution which is uniquely defined in the sector ${\cal S}_0$, where the sector ${\cal S}_k$ ($k\in {\mathbb Z}$) is defined by
\begin{equation}
   \mathcal{S}_k = \qty{ x\in\mathbb{C};
   \abs{ \arg (x) - \frac{2\pi  k}{N+h+1}}<\frac{\pi}{N+h+1}}.
\end{equation}
Using the Symanzik rotation, we are able to find the subdominant solution in ${\cal S}_k$:
\begin{equation}
    \phi_k(x,u_0,u_{N+1};\epsilon)=\phi(x,u_0,u_{N+1};e^{k\frac{2\pi i}{h}}\epsilon).
\end{equation}
Since both $\phi_k(e^{-2\pi i}x,u_0,u_N;\epsilon)$ and $\phi_{k+N+h+1}(x,u_0,u_N;\epsilon)$ are the subdominant solution in the same sector and the monodromy of the solution around the origin is trivial, the relation $\phi_k(e^{-2\pi i}x,u_0,u_N;\epsilon)\propto \phi_{k+N+h+1}(x,u_0,u_N;\epsilon)$ holds.
We thus find
\begin{equation}
    \phi_{k+N+h+1}(x,u_0,u_{N+1};\epsilon) \propto \phi_k(x,u_0,u_{N+1};\epsilon).
    \label{eq:periodicity_of_Symanzik_ro}
\end{equation}
Introduce the Wronskian of functions $f_i(x)$ $(i=0,\dots,r)$ by
\begin{equation}
    W\qty[f_{0},\dots,f_{r}] \coloneqq
    \det 
    \begin{pmatrix}
    f_{0} & \dots & f_{r}\\
    \vdots & \ddots & \vdots \\
    \partial_x^{r}f_{0} & \dots & \partial_x^{r}f_{r}
    \end{pmatrix}.
\end{equation}
One finds $W\qty[\phi_k,\dots,\phi_{k+r}] = 1$ due to the normalization factor, which implies the set of the subdominant solutions $\{\phi_k,\dots,\phi_{k+r}\}$ form a basis of the ODE.
Another important property of the Wronskian is
\begin{equation}
    W\qty[\phi_{i_0},\dots,\phi_{i_r}]^{[2l]} = W\qty[\phi_{i_0+l},\dots,\phi_{i_r+l}],\quad l\in \mathbb{Z},
    \label{eq:Wronskian_property}
\end{equation}
where we have used the notation
\begin{equation}
    g^{[l]}(u_0,u_N;\epsilon)\coloneqq g(u_0,u_N;e^{\frac{\pi il}{h}}\epsilon).
    \label{eq:Symanzik_rotation_of_epsilon}
\end{equation}
Now let us introduce the T-functions $T_{a,l}$ $(1\leq a\leq r,l\in\mathbb{Z})$ \cite{Ito:2017ypt}:
\begin{equation}
    \begin{aligned}
        &T_{a,l}\\
        =&\begin{dcases}
        W\qty[ \phi_{-r+1+\frac{a}{2}},\phi_{-r+2+\frac{a}{2}},\dots,\phi_{1-\frac{a}{2}},\phi_{l+2-\frac{a}{2}},\phi_{l+3-\frac{a}{2}},\dots,\phi_{l+1+\frac{a}{2}}]^{[-l-1]}, & a\text{: even},\\
        W\qty[\phi_{-r+1+\frac{a-1}{2}},\phi_{-r+2+\frac{a-1}{2}},\dots,\phi_{-\frac{a-1}{2}},\phi_{l+1-\frac{a-1}{2}},\phi_{l+2-\frac{a-1}{2}},\dots,\phi_{l+1+\frac{a-1}{2}}]^{[-l]}, & a\text{: odd}.
        \end{dcases}
    \end{aligned}
    \label{eq:def_T_system}
\end{equation}
Using the Pl\"ucker relation: 
\begin{equation}
\begin{aligned}
    W\qty[f_{0},\dots,f_{r-1},f_{r}]W&\qty[f_{0},\dots,f_{r-2},f_{r+1},f_{r+2}]\\
    &\qquad=W\qty[f_{0},\dots,f_{r-1},f_{r+1}]W\qty[f_{0},\dots,f_{r-2},f_{r},f_{r+2}]\\
    &\qquad\qquad +W\qty[f_{0},\dots,f_{r-1},f_{r+2}]W\qty[f_{0},\dots,f_{r-2},f_{r+1},f_{r}],
\end{aligned}
\end{equation}
and the property of the Wronskian \eqref{eq:Wronskian_property}, one can find that $T_{a,l}$ satisfies the functional relations called the T-system:
\begin{equation}
    T_{a,l}^{[+1]}T_{a,l}^{[-1]}=T_{a,l+1}T_{a,l-1} + T_{a+1,l}T_{a-1,l},
    \label{eq:Ar_T_system}
\end{equation}
where $T_{0,l}=T_{r+1,l}=1$. 
From \eqref{eq:def_T_system}, the T-functions satisfy the boundary conditions:
\begin{equation}
    T_{a,-1}=T_{a,N+2}=0,\qquad T_{a,0}=T_{a,N+1}=1,\qquad a=1,\dots,r.
\end{equation}
Consequently, the non-trivial T-functions are $T_{a,l}$ $(1\leq l\leq N)$.
Let us define the Y-functions from cross ratio of the T-functions by
\begin{equation}
    Y_{a,l}=\frac{T_{a-1,l}T_{a+1,l}}{T_{a,l+1}T_{a,l-1}},\qquad a=1,\dots,r,\qquad l\in\mathbb{Z}.
    \label{eq:def_Y_function}
\end{equation}
One can show the Y-functions satisfies the functional relations called the Y-system:
\begin{equation}
    Y_{a,l}^{[+1]}Y_{a,l}^{[-1]} = \frac{(1+Y_{a+1,l})(1+Y_{a-1,l})}{(1+Y_{a,l+1}^{-1})(1+Y_{a,l-1}^{-1})}.
    \label{eq:Ar_Y_system}
\end{equation}
Here we also defined $Y_{0,l}=Y_{r+1,l}=0$.
The boundary conditions for the Y-functions are given by
\begin{equation}
    Y_{a,0}^{-1}=Y_{a,N+1}^{-1}=0,\qquad a=1,\dots,r.
    \label{eq:boundary_condition_of_Ar_Y-system}
\end{equation}
We thus obtain the $(A_r, A_{N})$-type Y-system \cite{Zamolodchikov:1991et,Ravanini:1992fi}. 
Note that the Y-system is a universal concept associated with the Thermodynamic Bethe ansatz (TBA) equation \cite{Kuniba:1993cn}.
Fixing the asymptotic behavior of the Y-functions, it is straightforward to convert the Y-system into a unique set of TBA equations.
In the following subsection, we will examine the asymptotic behavior of the Y-function by using the WKB approximation and then compare the Y-functions with the WKB periods defined in the previous section.

\subsection{Y-functions and WKB periods}
\label{sec:Y_functions_as_periods}
In the previous subsection, we have defined the Y-functions as the cross ratio \eqref{eq:def_Y_function} of the Wronskians of the subdominant solutions.
We now discuss the leading term of the Y-functions in the WKB approximation.
The WKB solution of $\phi_k$ has the form of 
\begin{equation}
  \phi_k= \exp\qty[\frac{\delta_{k}}{e^{\frac{2\pi i}{h}k}\epsilon}\int_{x_{k}}^x S(x^{\prime};e^{\frac{2\pi i}{h}k}\epsilon) \dd x^{\prime}],
\end{equation}
where $x_{k}$ is the initial point of the integration.
$\delta_{k}$ is a phase factor, whose value depends on the sheets of the WKB curve in which $\phi_{k}$ lives.
To evaluate the Wronskian, one needs to choose $\delta_{k}$  and put $x$ as a common point, from which the Wronskian results in a $x$-independent function.
Performing this evaluation for the Wronskians in the Y-function, one can identify the $x_{k}$ in different Wronskians and obtain a cycle eventually \cite{Gaiotto:2009hg,Alday:2010vh}.
This suggests that the identification between the Y-functions and the WKB periods.

In the case of the second order ODE with the $(N+1)$-th order polynomial potential, say the $(A_1, A_N)$-type ODE, the identification between the WKB periods and the Y-functions has been obtained in \cite{Ito:2017ypt,Ito:2018eon}.
In general, we can determine the $\gamma$ by using the abelianization
\cite{Neitzke:2017yos,Hollands:2019wbr,Dumas:2020zoz}.
In \cite{Neitzke:2017yos}, one finds
\begin{equation}
    Y_{1,1} = \exp\qty[ \frac{1}{\epsilon}\Pi_{\gamma_1}],
    \label{eq:Y11_as_period}
\end{equation}
for the third order ODE \eqref{eq:Ar_ODE} with the quadratic potential, i.e. $p(x)=u_0x^2+u_2$.
In a similar way, one can find a formula for $Y_{2,1}^{[-1]}$. By considering the Stokes graph associated with the shift $\epsilon\to e^{-\frac{\pi i}{3}}\epsilon$, we obtain
\begin{equation}
    Y_{2,1}^{[-1]} = \exp\qty[ \frac{1}{\epsilon}\Pi_{\gamma_1+\gamma_2}].
    \label{eq:Y21_as_period}
\end{equation}
Note that we have defined the WKB period as a formal expansion in $\epsilon$. 
In order to define the equality in \eqref{eq:Y11_as_period} and \eqref{eq:Y21_as_period}, we have to take the Borel resummation of the WKB periods.
The $(r+1)$-th order ODE with the quadratic potential, say $(A_r, A_1)$-type ODE, can be regarded as the quantum Seiberg-Witten curve of the $(A_r, A_1)$-type AD theory which is dual to the $(A_1, A_r)$-type AD theory. 
This duality can be seen also at the level of the Y-system.
The $(A_r, A_1)$-type Y-system \eqref{eq:Ar_Y_system} is dual to the $(A_1, A_r)$-type Y-system from the $(A_1, A_r)$-type ODE.
It is thus natural to propose the identification between the Y-functions and the WKB periods for the higher order ODE:
\begin{equation}
    Y_{a,1}^{[-a+1]} = \exp \qty[\frac{1}{\epsilon}\Pi_{\gamma_1+\dots+\gamma_a}], \qquad a=1,\dots,r.
    \label{eq:Y_function_as_period}
\end{equation}
This will be tested numerically in the next section. 
From \eqref{eq:Y_function_as_period}, one can evaluate the asymptotic behaviors of the Y-functions.
The classical periods are given by
\begin{equation}
    \begin{aligned}
       \Pi_{\gamma_1+\cdots+\gamma_{a}}^{(0)}&=
        -\oint_{\gamma_1+\cdots+\gamma_{a}}p(x)^{\frac{1}{h}}\dd x\\
        &=-2 e^{\frac{\pi i}{h}(4 + a)}\sin\qty(\frac{\pi a}{h}) u_0^{-\frac{1}{2}} u_2^{\frac{1}{h}+\frac{1}{2}}  B\qty(\frac{1}{2},1+\frac{1}{h}).
    \end{aligned}
\end{equation}
Note that the classical periods satisfy the following relation:
\begin{equation}
    2 \Pi_{\gamma_1+\dots+\gamma_a}^{(0)}\cos(\frac{\pi}{h}) = \sum_{b = 1}^{r}G_{ab}e^{\frac{\pi i}{h}(a-b)}\Pi_{\gamma_1+\dots+\gamma_b}^{(0)},
    \label{eq:Perron-Frobenius}
\end{equation}
where $G_{ab}$ is the incidence matrix of the $A_r$-type Lie algebra.
Then the leading order approximation of the Y-function $Y_{a,1}$ becomes
\begin{equation}
    \begin{aligned}
        Y_{a,1} = \exp\qty[\frac{e^{-\frac{\pi i}{h}(a-1)}}{\epsilon}\Pi_{\gamma_1+\dots+\gamma_a}^{(0)} + \cdots].
    \end{aligned}
    \label{eq:leading_approximation_of_Y_function}
\end{equation}
We can check that \eqref{eq:leading_approximation_of_Y_function} satisfies the Y-system \eqref{eq:Y-system_for_ArA1} at the leading order in $\epsilon$. 

In the next section, we test the relation \eqref{eq:Y_function_as_period} numerically by solving the TBA equations satisfied by the Y-functions.

\section{TBA equations and numerical test}
\label{sec:TBA}
In this section, to confirm the identification \eqref{eq:Y_function_as_period}, we compare the corresponding terms in the $\epsilon$-expansions.
We rewrite the Y-system into the integral equations called the thermodynamic Bethe ansatz equations with the help of the asymptotic conditions of the Y-functions studied in section \ref{sec:Y_functions_as_periods}.
We then expand the TBA equation at small $\epsilon$, and compare it against the expansion of the WKB period.

\subsection{TBA equations}
\label{sec:derivation_of_TBA}
We first convert the Y-system into a set of integral equations for the Y-functions called the TBA equations according to \cite{Zamolodchikov:1991et, Dorey:2007zx}.
We consider particularly the $(A_r, A_1)$-type Y-systems \eqref{eq:Ar_Y_system}, where the second index $l$ of $Y_{a,l}$ takes value $l=1$ only.
The Y-system is written as
\begin{equation}
    \frac{Y_{a, 1}^{[+1]}Y_{a, 1}^{[-1]}}{\prod_{b = 1}^{r}Y_{b, 1}^{G_{ab}}} = \prod_{b = 1}^{r}(1 + Y_{b, 1}^{-1})^{G_{ab}}.
    \label{eq:Y-system_for_ArA1}
\end{equation}
Here the matrix $G_{ab}$ denotes the incidence matrix of $A_r$, which is defined by $G_{ab} \coloneqq 2\delta_{ab} - C_{ab}$.
$C_{ab}$ is the Cartan matrix of $A_r$.
Introducing the spectral parameter $\theta$ by $\epsilon = e^{-\theta}$, the rotation of $\epsilon$ defined in \eqref{eq:Symanzik_rotation_of_epsilon} acts on the Y-function, which is regarded as the function of $\theta$, as
\begin{equation}
    Y_{a, 1}^{[\pm 1]}(\theta) = Y_{a, 1}\qty(\theta \mp \frac{\pi i}{h}).
\end{equation}
At the large and positive real $\theta$, $\log{Y_{a, 1}}$ is assumed to behave as:
\begin{equation}
    \log{Y_{a, 1}(\theta)} \sim m_{a, 1}e^{\theta}.
\end{equation}
Here $m_{a, 1}$ is the mass parameter of the pseudo particle described by the TBA system.
Based on the observation in the previous section, the mass parameter is related to $\Pi_{\gamma_1 + \dots + \gamma_a}^{(0)}$ as
\begin{equation}
    m_{a, 1} = e^{-\frac{\pi i}{h}(a - 1)}\Pi_{\gamma_1 + \dots + \gamma_a}^{(0)}.
    \label{eq:mass1}
\end{equation}
Using this relations \eqref{eq:mass1} and \eqref{eq:Perron-Frobenius}, one can express $m_{a, 1}$ in terms of $m_{1,1}$ as
\begin{equation}
    m_{a, 1} = \frac{\sin(\frac{\pi a}{h})}{\sin(\frac{\pi}{h})}m_{1, 1}\;.
    \label{eq:mass_relation}
\end{equation}
This relation is essential to derive the TBA equations.
We take the logarithm of  the Y-system \eqref{eq:Y-system_for_ArA1} and express it in terms of $f_{a, 1} \coloneqq \log{Y_{a, 1}} - m_{a, 1}e^{\theta}$.
Taking the Fourier transform defined by
\begin{equation}
    \widetilde{f}(k) = \int_{-\infty}^{\infty}\dd{\theta}f(\theta)e^{-ik\theta},
\end{equation}
we obtain
\begin{equation}
    \sum_{b = 1}^{r}\qty(2\delta_{ab}\cosh(\frac{\pi k}{h}) - G_{ab})\widetilde{f}_{b, 1}(k) = \sum_{b = 1}^{r}G_{ab}\widetilde{L}_{b, 1}(k),
    \label{eq:tbaf1}
\end{equation}
where $L_{a, 1}$ is given by
\begin{equation}
    L_{a, 1}(\theta) \coloneqq \log(1 + Y_{a, 1}^{-1}(\theta)), \qquad L_{a, 0} = L_{a, 2} = 0.
\end{equation}
Solving \eqref{eq:tbaf1} in terms of $\widetilde{f}_{a, 1}(k)$ and performing the inverse Fourier transform, we finally obtain the TBA equations~\footnote{The TBA equations based on more general assumptions were derived in \cite{Suzuki:1999hu}. Fixing the asymptotic behaviors by using the WKB periods, the TBA equations \cite{Suzuki:1999hu} reproduce \eqref{eq:TBA_equations}.}:
\begin{equation}
    \log{Y_{a, 1}}(\theta) = m_{a, 1}e^{\theta} - \frac{1}{2\pi}\sum_{b = 1}^{r}\int_{-\infty}^{\infty}\mathcal{K}_{ab}(\theta - \theta^\prime)L_{b, 1}(\theta^\prime)\dd{\theta^\prime}.
    \label{eq:TBA_equations}
\end{equation}
The kernel $\mathcal{K}_{ab}(\theta)$ in the convolution term is defined by
\begin{equation}
    \mathcal{K}_{ab}(\theta) \coloneqq -\int_{-\infty}^{\infty}\qty[\sum_{c = 1}^{r}\qty(2\delta_{ac}\cosh(\frac{\pi k}{h}) - G_{ac})^{-1}G_{cb}]e^{ik\theta}\dd{k},
\end{equation}
which turns out to be \cite{Dorey:2007zx}
\begin{equation}
    \mathcal{K}_{ab}(\theta) = -i\dv{\theta}\sum_{\substack{x = \abs{a - b} + 1 \\ \mathrm{step} 2}}^{a + b - 1}\log\qty{x},
    \label{eq:kernel_function}
\end{equation}
where the functions $\qty{x}$ is given by
\begin{equation}
    \qty{x} \coloneqq \frac{\sinh(\frac{\theta}{2} + \frac{\pi i}{2h}(x - 1))}{\sinh(\frac{\theta}{2} - \frac{\pi i}{2h}(x - 1))}\frac{\sinh(\frac{\theta}{2} + \frac{\pi i}{2h}(x + 1))}{\sinh(\frac{\theta}{2} - \frac{\pi i}{2h}(x + 1))}.
\end{equation}
The TBA equations \eqref{eq:TBA_equations} have the same form as the ones in \cite{Dorey:2007zx}, but can have complex valued masses depending on $u_0$ and $u_2$.
In a region of parameters $u_0$ and $u_2$ such that the real part of $m_{a,1}$ is positive, the Y-function  calculated from \eqref{eq:TBA_equations} is shown to converge numerically.
On the other hand, the Y-function diverges numerically when the real part of $m_{a, 1}$ is negative.
To resolve this problem, we shift $\theta$ to $\theta-i\phi$ such that  ${\rm Re}(m_{a, 1}e^{-i\phi})$ becomes positive.

From the definition of the kernel function of the TBA equations \eqref{eq:kernel_function}, the $a$-th and $(r + 1 - a)$-th equation are the same, and together with the mass relation \eqref{eq:Perron-Frobenius}, we find that the Y-function has the following symmetries:
\begin{equation}
    Y_{a, 1} = Y_{r + 1 - a, 1}, \qquad a = 1, \dots, r.
    \label{eq:symmetry_of_Y-function}
\end{equation}

We investigate the correspondence between the period integral and the Y-function in the $\epsilon$-expansion.
To see this, we perform the $e^{-\theta}$-expansion of the kernel function of the TBA equations following \cite{Zamolodchikov:1989cf}.
Using the expansion for the functions $\qty{x}$, one finds
\begin{equation}
    -i\dv{\theta}\ln\qty{x} = -4\sum_{n = 1}^{\infty}\cos(\frac{\pi n}{h})\sin(\frac{\pi n}{h}x)e^{-n\theta}.
\end{equation}
For $a \leq b$, we can expand the kernel functions in $e^{-\theta}$ as
\begin{equation}
    \begin{aligned}
        \mathcal{K}_{ab}(\theta) &= -4\sum_{n = 1}^{\infty}\cos(\frac{\pi n}{h})\sum_{j = 1}^{a}\sin(\frac{\pi n(b - a - 1 + 2j)}{h})e^{-n\theta}\\
        &= -4\sum_{\substack{n = 1 \\ n \neq 0 \mod h}}^{\infty}\cot(\frac{\pi n}{h})\sin(\frac{\pi n a}{h})\sin(\frac{\pi n b}{h})e^{-n\theta}.
    \end{aligned}
\end{equation}
Note that the terms $e^{-nh\theta}$ $(n=1,2,\dots)$ are absent in the expansion.
Finally, the Y-functions have the following asymptotic expansion in $e^{-\theta}$:
\begin{equation}
    \log{Y_{a, 1}(\theta)} = m_{a, 1}e^{\theta} + \sum_{n = 1}^{\infty}m_{a, 1}^{(n)}e^{-n\theta},
    \label{eq:expansion_of_log_Y}
\end{equation}
where the coefficients $m_{a, 1}^{(n)}$ are given by
\begin{equation}
    m_{a, 1}^{(n)} \coloneqq
    \begin{dcases}
        \frac{2}{\pi}\cot(\frac{\pi n}{h})\sin(\frac{\pi n a}{h})\sum_{b = 1}^{r}\sin(\frac{\pi n b}{h})\int_{-\infty}^{\infty}L_{b, 1}(\theta')e^{n\theta'}\dd{\theta'}, & n \neq 0 \mod h,\\
        0, & n = 0 \mod h.
    \end{dcases}
    \label{eq:correction_of_m}
\end{equation}
In the following subsection, we will confirm that these coefficients $m_{a, 1}^{(n - 1)}$ correspond to period integrals $\Pi^{(n)}_{\gamma_1 + \dots + \gamma_a}$.

\subsection{Numerical test}
\label{sec:numerical_comparison}
In this subsection, we compare the WKB period with the Y-function numerically and confirm the identification \eqref{eq:Y_function_as_period}:
\begin{equation}
    \log{Y_{a, 1}} = \qty(\frac{1}{\epsilon}\Pi_{\gamma_1 + \cdots \gamma_a})^{[a-1]}.
    \label{eq:ide1}
\end{equation}
Then (\ref{eq:ide1}) implies
\begin{equation}
    m_{a, 1}^{(n - 1)} = e^{\frac{\pi i}{h}(a-1)(n-1)}\Pi^{(n)}_{\gamma_1 + \cdots + \gamma_a}, \qquad n = 0,1,\dots,
    \label{eq:coeff1}
\end{equation}
where we have defined $m_{a,1}^{(-1)}=m_{a,1}$.
For $n-1=k h$ $(k=1,2,\dots)$, we have seen that the l.h.s. of (\ref{eq:coeff1}) vanishes.
For the WKB analysis, we also see that the corresponding term in the r.h.s. also vanishes.
For example, for even $r$ and $k=2i+1$, we obtain $n=2h i+h+1$ that we have seen in section \ref{sec:WKB period}.

We now check the equation (\ref{eq:coeff1}) numerically.
To solve the TBA equations, we use the fast Fourier transformation (FFT) with $2^{20}$ discrete points and the cutoff 16.
The results for $A_2$ and $A_3$ case with $p(x) = x^2 - 1$ are shown in Tables \ref{tab:A2A1_x^2+1} and \ref{tab:A3A1_x^2+1}, respectively.
We can see that the two results match each other with high precision.

As another consistency check of (\ref{eq:ide1}), we investigate the ${\mathbb Z}_2$-symmetry of the WKB periods.
As we mention in the section \ref{sec:derivation_of_TBA}, the Y-functions of the $(r + 1)$-th order ODE have the symmetry \eqref{eq:symmetry_of_Y-function}.
From the identification \eqref{eq:ide1}, for the third order ODE, the WKB periods should also have this symmetry:
\begin{equation}
    \epsilon^{-1}\Pi_{\gamma_1} = \qty(\epsilon^{-1} \Pi_{\gamma_1+\gamma_2})^{[+1]}.
    \label{eq:sym_pi}
\end{equation}
We can check the above equation explicitly.
In fact, the periods
$\epsilon^{-1}\Pi_{\gamma_1}$ and $\qty(\epsilon^{-1}\Pi_{\gamma_1+\gamma_2})^{[+1]}$ are given by
\begin{align}
    \epsilon^{-1}\Pi_{\gamma_1} 
    &= \sum_{n\geq0}\qty( \epsilon^{6n-1} a_{6n} \hat{\Pi}_{\gamma_1}^{(0),1} + \epsilon^{6n+1} a_{6n+2} \hat{\Pi}_{\gamma_1}^{(0),2}),\\
    \qty(\epsilon^{-1} \Pi_{\gamma_1+\gamma_2})^{[+1]} 
    &= \sum_{n\geq0}\qty( \epsilon^{6n-1} a_{6n} e^{-\frac{\pi i}{3}}\hat{\Pi}_{\gamma_1+\gamma_2}^{(0),1} + \epsilon^{6n+1} a_{6n+2} e^{\frac{\pi i}{3}} \hat{\Pi}_{\gamma_1+\gamma_2}^{(0),2}),
\end{align}
where we defined $a_n = \Pi_{\gamma}^{(n)}/\hat{\Pi}_{\gamma}^{(0),l_n}$.
Since 
\begin{equation}
     \hat{\Pi}_{\gamma_1}^{(0),1} = e^{-\frac{\pi i}{3}}{\hat{\Pi}_{\gamma_1+\gamma_2}^{(0),1}}, \qquad \hat{\Pi}_{\gamma_1}^{(0),2} = e^{\frac{\pi i}{3}} {\hat{\Pi}_{\gamma_1+\gamma_2}^{(0),2}},
\end{equation}
one finds \eqref{eq:sym_pi}.
For the higher order ODE, one can also check the relation similarly.

We can also use the equation \eqref{eq:ide1} for the third order ODE to compare the eigenvalues of the integral of motions in the $W_3$ conformal field theory with the expansion of the T-functions \cite{Bazhanov:2001xm}.
In appendix \ref{sec:T-expansion}, we show the identification between the large spectral parameter expansion of their T-function and the $\epsilon$-expansion of our WKB period.

\begin{table}[H]
    \centering
    \small
    \begin{tabular}{r||r|r}
        $n$ & $\Pi^{(n)}_{\gamma_1},e^{\frac{\pi i}{3}(n-1)}\Pi^{(n)}_{\gamma_1+\gamma_2}$ & $m_{1, 1}^{(n - 1)},m_{2, 1}^{(n - 1)}$\\
        \hline
        $2$ & $0.1244723667 i$ & $0.1244723666i$\\
        $4$ & $0$ & $0$\\
        $6$ & $-0.05721560699 i$ & $-0.05721561024i$\\
        $8$ & $-0.2089662087 i$ & $-0.2089662250i$\\
        $10$ & $0$ & $0$\\
        $12$ & $14.97696460 i$ & $14.97696419i$\\
        $14$ & $234.1765144 i$ & $234.1761561i$
    \end{tabular}
    \caption{$\Pi_{\gamma}^{(n)}$ and $m_{a,1}^{(n-1)}$ for the third order ODE with $p(x) = x^2 - 1$.}
    \label{tab:A2A1_x^2+1}
\end{table}

\begin{table}[H]
    \centering
    \small
    \begin{tabular}{r||r|r||r|r}
        $n$ & $\Pi^{(n)}_{\gamma_1},e^{\frac{\pi i}{2}(n-1)}\Pi^{(n)}_{\gamma_1+\gamma_2+\gamma_3}$ & $m_{1, 1}^{(n - 1)},m_{3, 1}^{(n - 1)}$ & $e^{\frac{\pi i}{4}(n-1)}\Pi^{(n)}_{\gamma_1 + \gamma_2}$ & $m_{2, 1}^{(n - 1)}$\\
        \hline
        $2$ & $-0.2118032712$ & $-0.2118032752$ & $-0.2995350587$ & $-0.2995350643$\\
        $4$ & $0.05311151419$ & $0.05311151743$ & $-0.07511102369$ & $-0.07511102827$\\
        $6$ & $-0.12953645375$ & $-0.1295364668$ & $0.1831922097$ & $0.1831922283$\\
        $8$ & $0.7882359521$ & $0.7882360637$ & $1.1147339738$ & $1.114734131$\\
        $10$ & $-7.184548229$ & $-7.184549472$ & $-10.160485545$ & $-10.16048730$\\
        $12$ & $102.58179442$ & $102.5818017$ & $-145.0725649+$ & $-145.0725752$\\
        $14$ & $-2251.106503$ & $-2251.103294$ & $3183.545348$ & $3183.540809$
    \end{tabular}
    \caption{$\Pi_{\gamma}^{(n)}$ and $m_{a,1}^{(n-1)}$ for the fourth order ODE with $p(x) = x^2 - 1$.}
    \label{tab:A3A1_x^2+1}
\end{table}

\subsection{PNP relation}
We discuss the relation among the WKB periods, which is obtained as a consequence of the identification in \eqref{eq:coeff1}.
The effective central charge associated to our TBA equations is given by
\begin{equation}
    c_{\mathrm{eff}} \coloneqq \frac{3}{\pi^2}\sum_{a = 1}^{r}\int_{-\infty}^{\infty}m_{a, 1}L_{a, 1}(\theta)e^{\theta}\dd{\theta}.
\end{equation}
Using the Roger's dilogarithm identity, the effective central charge becomes
\begin{equation}
    c_{\mathrm{eff}} = \frac{h - 1}{h + 2}.
\end{equation}
By computing $m_{a, 1}^{(1)}$ from the definition \eqref{eq:correction_of_m} and using the identification between the mass and the period \eqref{eq:coeff1}, we obtain the following relation:
\begin{equation}
    \Pi_{\gamma_1 + \dots + \gamma_a}^{(0)}\Pi_{\gamma_1 + \dots + \gamma_a}^{(2)} = \frac{2\pi}{3}\cot(\frac{\pi}{h})\sin[2](\frac{\pi a}{h})c_{\mathrm{eff}} = \frac{2\pi}{3}\frac{h - 1}{h + 2}\cot(\frac{\pi}{h})\sin[2](\frac{\pi a}{h}).
    \label{eq:PNP_rel_02}
\end{equation}
Here we do not sum over $a$.
This relation is regarded as the PNP relation \cite{Dunne:2014bca,Codesido:2016dld}.
The relation (\ref{eq:PNP_rel_02}) can also be obtained directly from the calculation of the periods.
The classical period and the second order correction are given by
\begin{equation}
    \begin{aligned}
        \Pi_{\gamma_1 + \dots + \gamma_a}^{(0)} &= -2 e^{\frac{\pi i}{h}(4 + a)}\sin(\frac{\pi a}{h}) u_0^{\frac{1}{h} + \frac{1}{2}} u_2^{-\frac{1}{2}} B\qty(\frac{1}{2}, 1 + \frac{1}{h}),\\
        \Pi_{\gamma_1 + \dots + \gamma_a}^{(2)} &=- \frac{4r^2 - h^2}{24h}e^{-\frac{\pi i}{h}(4+a)}\sin(\frac{\pi a}{h})  u_0^{\frac{r}{h}-\frac{3}{2}} u_2^{\frac{1}{2}}B\qty(\frac{1}{2},1+\frac{r}{h})
    \end{aligned}
\end{equation}
Then their product becomes
\begin{equation}
    \begin{aligned}
        \Pi_{\gamma_1 + \dots + \gamma_a}^{(0)}\Pi_{\gamma_1 + \dots + \gamma_a}^{(2)} &= \frac{h}{12}4\sin^2\qty(\frac{\pi a}{h})\frac{\Gamma\qty(\frac{1}{2})^2\Gamma\qty(1+\frac{1}{h})\Gamma\qty(1+\frac{r}{h})}{\Gamma\qty(\frac{3}{2}+\frac{1}{h})\Gamma\qty(-\frac{1}{2}+\frac{r}{h})}.
    \end{aligned}
\end{equation}
By using the reflection formula for the gamma function $\Gamma(z)\Gamma(1-z) = \pi/\sin(\pi z)$, one obtains \eqref{eq:PNP_rel_02}.

\section{Higher order potential and Wall-crossing %of TBA equations
}\label{sec:wall_crossing}
So far, we have studied the higher order ODE with a monomial potential.
It is interesting to study polynomial potential, where the WKB periods and the TBA equations show the wall-crossing phenomena\cite{Gaiotto:2009hg}.
For the second order ODE, the wall-crossing of the TBA equations and the WKB periods has been studied in \cite{Ito:2018eon}.
In this section, we observe that the wall-crossing phenomena occur by considering the third order ODE with a cubic potential $p(x)=u_0x^3+u_1x^2+u_2x+u_3$ as an example. 

Following the procedure of section 3, it is easy to generalize the Y-system (\ref{eq:Ar_Y_system}) to $(A_r,A_N)$ type with Y-functions $Y_{a,s}$ ($a=1,\dots,r$, $s=1,\dots, N$), which 
leads to the $(A_r,A_N)$-type TBA equations \cite{Suzuki:1999hu}. 
In the minimal chamber of $(A_2,A_2)$-type Y-system, they can be written as
\begin{equation}
    \begin{aligned}
    \label{A2A2-TBA-min}
       \log Y_{a,1}(\theta-i\phi_{1})=&|m_{a,1}|e^{\theta}+K\star\overline{L}_{a,1}-K_{1,2}\star\overline{L}_{a,2}\\
       \log Y_{a,2}(\theta-i\phi_{2})=&|m_{a,2}|e^{\theta}-K_{2,1}\star\overline{L}_{a,1}+K\star\overline{L}_{a,2},\quad (a=1,2)
    \end{aligned}
\end{equation}
where $K(\theta)=\frac{1}{2\pi}\frac{4\sqrt{3}\cosh{\theta}}{1 + 2\cosh{2\theta}}$, $\phi_s={\rm Arg}(m_{1,s})$ and $\overline{L}_{1,s}(\theta)=\log\big(1+Y_{1,s}(\theta-i\phi_s)^{-1}\big)$.
Here
$K_{r,s}$ denotes  the shifted kernel which is defined by
\begin{equation}
K_{r,s}(\theta)= K(\theta-i\phi_{r}+i\phi_{s}).
\end{equation}
Since the TBA equations for $a=2$ provide the same copy as $a=1$. We only consider the $a=1$ equations.

By using the WKB approximations and the Stokes graph in \cite{Neitzke:2017yos}, one can compute the leading order contribution to the Y-functions at large $e^\theta$ as in (\ref{eq:leading_approximation_of_Y_function}). We then find the masses and the classical periods are related by
\begin{equation}
  m_{a,s}=e^{\frac{\pi i}{3}(s-a)}\Pi^{(0)}_{\hat{\gamma}_{a,s}} ,\qquad a,s=1,2,
    \label{eq:mass_as_Pi}
\end{equation}
where $\Pi^{(0)}_{\hat{\gamma}_{a,s}}$ denotes the classical SW period for the 1-cycle
$\hat{\gamma}_{a,s}$. See  Fig.\ref{fig:cycle-wc} (a) for an example of $\hat{\gamma}_{a,s}$. Including higher order contributions in $\epsilon=e^{-\theta}$, we propose that the Y-functions are identified to the exact WKB periods 
\begin{equation}
    \log Y_{1,1}(\theta)=\frac{1}{\epsilon}\Pi_{\hat{\gamma}_{1,1}}(\theta), \quad \log Y_{1,2}(\theta)=\Big(\frac{1}{\epsilon}\Pi_{\hat{\gamma}_{1,2}}\Big)(\theta+\frac{\pi i}{3}),
\end{equation}
which will be tested numerically. At large $e^{\theta}$, expanding the kernels in  the TBA equations, the Y-functions are expressed as
\begin{equation}
    \log{Y_{a, s}}(\theta) = m_{a, s} e^\theta + \sum_{n = 1}^{\infty}m_{a, s}^{(n)}e^{-n\theta}.
    \label{eq:y-exp1}
\end{equation}
The coefficients $m_{a, s}^{(n)}$ are given by
\begin{equation}
    m_{a, s}^{(n)} = k_n\int_{-\infty}^{\infty}(\overline{L}_{a, s}(\theta)e^{n(\theta - i\phi_{a, s})} - \overline{L}_{a,s - 1}(\theta)e^{n(\theta - i\phi_{a, s - 1})} - \overline{L}_{a, s + 1}(\theta)e^{n(\theta - i\phi_{a, s + 1})})\dd{\theta},
\end{equation}
where $\overline{L}_{a,0}=\overline{L}_{a,3}=0$ and the coefficient $k_n$ is defined by
\begin{equation}
    k_n = \frac{1}{\pi}\qty(\sin(\frac{\pi}{3}n) + \sin(\frac{2\pi}{3}n)).
    \label{eq:m-coefficient}
\end{equation}
Moreover, the higher order correction of WKB periods can be computed by using the Picard-Fuchs operators \eqref{eq:def_PF_operator}. We now confirm this identification numerically by comparing the expansions of WKB periods and Y-functions. 
In Table \ref{tab:A2A2_min}, we computed the corrections to the quantum periods and the coefficients of the expansion \eqref{eq:y-exp1} for the WKB curve $y^3=-x^3+x^2+2x$ which has branch points at $x=-1,0$ and $2$. The classical SW periods are defined as
\begin{equation}
    \Pi^{(0)}_{\hat{\gamma}_{1,1}}=\int_{\hat{\gamma}_{1,1}}y\dd{x},\qquad
    \Pi^{(0)}_{\hat{\gamma}_{1,2}}=\int_{\hat{\gamma}_{1,2}}y\dd{x}.
\end{equation}
which are calculated by the hypergeometric integral. The masses and the classical periods are related by \eqref{eq:mass_as_Pi}.
From Table \ref{tab:A2A2_min}, one finds that two results agree with each other numerically.

\begin{table}[H]
    \centering
    \small
    \begin{tabular}{r||r|r||r|r}
        $n$ & $\Pi_{\hat{\gamma}_{1,1}}^{(n)}$ & $m_{1, 1}^{(n - 1)}$ & $e^{-\frac{\pi i}{3}(n-1)}\Pi_{\hat{\gamma}_{1,2}}^{(n)}$ & $m_{1, 2}^{(n - 1)}$ \\
        \hline
        $0$ & $3.642563830i$ & $3.642563830i$ & $1.244358075i$ & $1.244358075i$ \\
        $2$ & $0.1994669082i$ & $0.1994669082i$ & $-0.1994669082i$ & $-0.1994669082i$ \\
        $4$ & $0$ & $0$ & $0$ & $0$ \\
        $6$ & $-4.120900959i$ & $-4.120900963i$ & $4.120900959i$ & $4.120900963i$ \\
        $8$ & $-82.25658234i$ & $-82.25658241i$ & $82.25658234i$ & $82.25658241i$ \\
        $10$ & $0$ & $0$ & $0$ & $0$ \\
        $12$ & $176192.6118i$ & $176192.5854i$ & $-176192.6118i$ & $-176192.5854i$ \\
        $14$ & $15083126.44i$ & $15083101.15i$ & $-15083126.44i$ & $-15083101.15i$
    \end{tabular}
    \caption{$\Pi_{\hat{\gamma}_{1,s}}^{(n)}$ and $m_{1,s}^{(n-1)}$ $(s=1,2)$ for the third order ODE with $p(x) = -x^3+x^2+2x$. The branch points are $x_0 = 2$, $x_1 = 0$, $x_2 = -1$. Here we have denoted the mass term $m_{a,s}$ by $m_{a,s}^{(-1)}$.}
    \label{tab:A2A2_min}
\end{table}
The kernel $K_{1,2}$ is shown to have poles on the real axis when $|\phi_2-\phi_1|=\frac{\pi}{3},\frac{2\pi}{3},\cdots$. 
Comparing with the quadratic potential, the phase of the mass parameters are not necessarily the same and the integration path of the TBA equations could cross the pole, where the wall-crossing phenomena occur \cite{Ito:2018eon}. The TBA equations (\ref{A2A2-TBA-min}) are valid in the minimal chamber, namely the region $|\phi_2-\phi_1|<\frac{\pi}{3}$. 
When $\phi_2-\phi_{1}$ crosses $\pm \frac{\pi}{3}, \pm \frac{2\pi}{3},\cdots$, one needs to modify the TBA equations by picking the contributions of the poles. This modification of TBA equations is known as the wall-crossing \cite{Alday:2010vh,Toledo}.

Let us consider the situation where $\phi_2-\phi_1$ crosses $\pi/3$. We thus need to pick the residue of pole in $K$ in the TBA equations:
\begin{equation}
\label{2-TBA}
    \begin{aligned}
    \log Y_{1,1}(\theta-i\phi_{1})=&|m_{1,1}|e^{\theta}+K\star\overline{L}_{1,1}-K_{1,2}\star\overline{L}_{1,2}-L_{1,2}(\theta-\frac{\pi i}{3}-i\phi_{1}),\\
    \log Y_{1,2}(\theta-i\phi_{2})=&|m_{1,2}|e^{\theta}-K_{2,1}\star\overline{L}_{1,1}+K\star\overline{L}_{1,2}-L_{1,1}(\theta+\frac{\pi i}{3}-i\phi_{2}).
   \end{aligned}
\end{equation}
To obtain a closed system, we also need to shift the spectral parameter of $Y_{1,1}$ and $Y_{1,2}$ to obtain the equations for $\log Y_{1,1}(\theta+\frac{\pi i}{3}-i\phi_2)$ and $\log Y_{1,2}(\theta-\frac{\pi i}{3}-i\phi_1)$. We thus obtain a closed system with four TBA equations.

It would be more interesting to introduce a set of new Y-functions $Y^{\rm n}$:
\begin{equation}
\begin{aligned}
Y_{1,1}^{\rm n}(\theta)&=Y_{1,1}(\theta)\big(1+\frac{1}{Y_{1,2}(\theta-\frac{\pi i}{3})}\big), \quad Y_{1,2}^{\rm n}(\theta)=Y_{1,2}(\theta)\big(1+\frac{1}{Y_{1,1}(\theta+\frac{\pi i}{3})}\big)\\
  Y_{12}^{\rm n}(\theta)&=\frac{1+\frac{1}{Y_{1,2}(\theta-\frac{\pi i}{3})}+\frac{1}{Y_{1,1}(\theta)}}{\frac{1}{Y_{1,1}(\theta)Y_{1,2}(\theta-\frac{\pi i}{3})}}=Y_{1,1}(\theta)Y_{1,2}(\theta-\frac{\pi i}{3})+Y_{1,1}(\theta)+Y_{1,2}(\theta-\frac{\pi i}{3})
\end{aligned}
\end{equation}
to absorb the residue in the right hand side of (\ref{2-TBA}). The TBA system in this case reads
\begin{align}
\log Y_{1,1}^{{\rm n}}(\theta-i\phi_{1})=&|m_{1,1}|e^{\theta}+K\star\overline{L}_{1,1}^{{\rm n}}-K_{1,2}\star\overline{L}_{1,2}^{{\rm n}}+K_{1,12}^{-}\star\overline{L}_{12}^{{\rm n}},\notag\\
\log Y_{1,2}^{{\rm n}}(\theta-i\phi_{2})=&|m_{1,2}|e^{\theta}+K\star\overline{L}_{1,2}^{{\rm n}}-K_{2,1}\star\overline{L}_{1,1}^{{\rm n}}-K_{2,12}^{-}\star\overline{L}_{12}^{{\rm n}},\label{eq:new-tba}\\
\log Y_{12}^{{\rm n}}(\theta-i\phi_{12})=&|m_{12}|e^{\theta}+K\star\overline{L}_{12}^{{\rm n}}+K_{12,1}^{+}\star\overline{L}_{1,1}^{{\rm n}}-K_{12,2}^{+}\star\overline{L}_{1,2}^{{\rm n}},\notag
\end{align}
where $K_{r,s}^{\pm}(\theta)=K_{r,s}(\theta\pm\frac{\pi i}{3})$, $\overline{L}^{\rm n}_{1,s}(\theta)=\log\big(1+Y^{\rm n}_{1,s}(\theta-i\phi_s)^{-1}\big)$ and $\overline{L}^{\rm n}_{12}(\theta)=\log\big(1+Y_{12}^{\rm n}(\theta-i\phi_{12})^{-1}\big)$. The mass term of $Y_{12}^{\rm n}$ is $m_{12}=m_{1,1}+e^{-\frac{\pi i}{3}}m_{1,2}$, whose phase is denoted by $\phi_{12}$.
From the asymptotic behavior of the TBA equations, we find the new Y-functions $Y_{1,1}^{\rm n}$ and $Y_{1,2}^{\rm n}$ are related to the cycles $\hat{\gamma}_{1,1}$ and $\hat{\gamma}_{1,2}$. Note the relation
\begin{equation}
    m_{12}=m_{1,1}+e^{-\frac{\pi i}{3}}m_{1,2}=\Pi^{(0)}_{\hat{\gamma}_{1,1}}+\Pi^{(0)}_{\hat{\gamma}_{1,2}},
\end{equation}
the $Y^{\rm n}_{12}$ is related with the cycle $\hat{\gamma}_{1,1}+\hat{\gamma}_{1,2}$. See Fig.\ref{fig:cycle-wc} (b). From the point view of the Argyres-Douglas theory, the BPS states in the minimal chamber are related with the cycles basis $(\hat{\gamma}_{1,1}, \hat{\gamma}_{1,2})$. In next chamber, the basis of BPS state charge becomes $(\hat{\gamma}_{1,1}, \hat{\gamma}_{1,2}, \hat{\gamma}_{1,1}+\hat{\gamma}_{1,2})$ \cite{Gaiotto:2009hg}. To study the full spectrum of the next chamber, one needs to consider the closed system related with the three cycles. Rather than the original Y-function, one thus needs to identify the new Y-function with the WKB periods. We thus propose that the new Y-functions are identified with the WKB periods at the region $\pi/3<\phi_2-\phi_1<2\pi/3$: 
\begin{align}
     \log Y_{1,1}^{\rm n}(\theta)&=\Big(\frac{1}{\epsilon}\Pi_{\hat{\gamma}_{1,1}}\Big)(\theta), \quad \log Y_{1,2}^{\rm n}(\theta)=\Big(\frac{1}{\epsilon}\Pi_{\hat{\gamma}_{1,2}}\Big)(\theta+\frac{\pi i}{3}),\nonumber\\
     \log Y^{\rm n}_{12}(\theta)&=\Big(\frac{1}{\epsilon}\Pi_{\hat{\gamma}_{1,1}}\Big)(\theta)
     +\Big(\frac{1}{\epsilon}\Pi_{\hat{\gamma}_{1,2}}\Big)(\theta)=:\Big({1\over \epsilon} \Pi_{\hat{\gamma}_{1,12}}\Big)(\theta),
     \label{eq:new-y-exp1}
\end{align}
where $\hat{\gamma}_{1,12}\coloneqq\hat{\gamma}_{1,1}+\hat{\gamma}_{1,2}$.
The large $\epsilon^{-1}=e^\theta$ expansion of (\ref{eq:new-y-exp1}) can be checked by solving the TBA equations numerically, which is shown in Tables \ref{tab:A2A2_next} and \ref{tab:A2A2_next_2}. Here we have defined $(m^{\rm n}_{a,k})^{(n)}$ and $(m^{\rm n}_{12})^{(n)}$ by the expansions: 
\begin{equation}
    \begin{aligned}
        \log{Y^{\rm n}_{a, k}}(\theta) &= m_{a,k} e^\theta + \sum_{n = 1}^{\infty}(m^{\rm n}_{a, k})^{(n)}e^{-n\theta}, \quad k=1,2,\\
        \log{Y^{\rm n}_{12}}(\theta) &= m_{12} e^\theta + \sum_{n = 1}^{\infty}(m^{\rm n}_{12})^{(n)}e^{-n\theta}.
    \end{aligned}
    \label{eq:y-exp2}
\end{equation}

\begin{figure}[htbp]
    \centering
    \begin{minipage}[b]{0.45\linewidth}
        \centering
        \begin{tikzpicture}
        \def\ra{1.7};
        \def\dv{0.07};
        \def\rt{1.5};
        \def\rv{0.2};
        \def\dd{0.05};
        \coordinate[] (ni) at (-3, 0);
        \coordinate[] (pi) at (3, 0);
        \coordinate[] (0i) at (-0.5, 1.5);
        \coordinate[label=above:] (x2) at ({-\rt}, 0);
        \coordinate[label=above:] (x1) at (-0.5, 0);
        \coordinate[label=above:] (x0) at ({\rt}, 0);
        \coordinate[] (x2p) at ({-\ra}, {\dv});
        \coordinate[] (x2n) at ({-\ra}, {-\dv});
        \coordinate[] (x0p) at ({\ra}, {\dv});
        \coordinate[] (x0n) at ({\ra}, {-\dv});
        \coordinate[] (x1p) at ($({\dv}, {\rv})+(-0.5, 0)$);
        \coordinate[] (x1n) at ($({-\dv}, {\rv})+(-0.5, 0)$);
        
        \coordinate[] (x2ip) at ({-\rt}, {\rv});
        \coordinate[] (x2in) at ({-\rt}, {-\rv});
        \coordinate[] (x0ip) at ({\rt}, {\rv});
        \coordinate[] (x0in) at ({\rt}, {-\rv});
        \coordinate[] (x1ip) at ($({\rv}, 0)+(-0.5, 0)$);
        \coordinate[] (x1in) at ($({-\rv}, 0)+(-0.5, 0)$);
        \coordinate[] (x1vn) at (-0.5, {-\rv});
        
        \fill (x2) circle [radius=2pt, red];\draw ({-\rt}, -0.3) node[below]{$-1$};
        \fill (x1) circle [radius=2pt, red];\draw (-0.5, -0.3) node[below]{$0$};
        \fill (x0) circle [radius=2pt, red];\draw ({\rt}, -0.3) node[below]{$2$};
        
        \draw[line width=0.4mm, gray, dashed] (ni) -- (x2);
        \draw[line width=0.4mm, gray, dashed] (x1) -- (0i);
        \draw[line width=0.4mm, gray, dashed] (x0) -- (pi);
        
        \draw[line width=0.3mm, red] (x2p) to [out=85, in=180] (x2ip);
        \draw[line width=0.3mm, red] (x2ip) to [out=0, in=180] ($(x1vn)-(0,{\dd})$);
        \draw[line width=0.3mm, red] ($(x1vn)-(0,{\dd})$) to [out=0, in=270] ($(x1ip)-(0,{\dd})$);
        \draw[line width=0.3mm, red] ($(x1ip)-(0,{\dd})$) to [out=90, in=-5] ($(x1p)-(0,{\dd})$);
        
        \draw[line width=0.3mm, dashed, red] ($(x1n)-(0,{\dd})$) to [out=185, in=0] (x2in);
        \draw[line width=0.3mm, dashed, red] (x2in) to [out=180, in=275] (x2n);

        \draw[line width=0.3mm, black] ($(x1p)+(0,{\dd})$) to [out=-5, in=180] (x0in);
        \draw[line width=0.3mm, black] (x0in) to [out=0, in=265] (x0n);
        
        \draw[line width=0.3mm, dashed, black] (x0p) to [out=95, in=0] (x0ip);
        \draw[line width=0.3mm, dashed, black] (x0ip) to [out=180, in=0] ($(x1vn)+(0,{\dd})$);
        \draw[line width=0.3mm, dashed, black] ($(x1vn)+(0,{\dd})$) to [out=180, in=270] ($(x1in)+(0,{\dd})$);
        \draw[line width=0.3mm, dashed, black] ($(x1in)+(0,{\dd})$) to [out=90, in=185] ($(x1n)+(0,{\dd})$);
        
        \end{tikzpicture}
        \subcaption{}
    \end{minipage}
    \begin{minipage}[b]{0.45\linewidth}
        \centering
        \begin{tikzpicture}
        \def\ra{1.7};
        \def\dv{0.03};
        \def\rt{1.5};
        \def\rv{0.2};
        \def\dd{0.04};
        \def\vsh{1.5};
        \def\hsh{0};
        \def\argAA{63.4};\def\argAB{-116.6};\def\argaa{26.6};\def\argab{153.4};
        \def\argBA{45};\def\argBB{-135};
        \coordinate[] (ni) at (-3, 0);
        \coordinate[] (pi) at ($(3, 0) + ({-\hsh}, {-\vsh})$);
        \coordinate[] (0i) at (0, 1);
        \coordinate[label=above:] (x2) at ({-\rt}, 0);
        \coordinate[label=above:] (x1) at (0, 0);
        \coordinate[label=above:] (x0) at ($({\rt}, 0) + ({-\hsh}, {-\vsh})$);
        \coordinate[] (x2p) at ({-\ra}, {\dv});
        \coordinate[] (x2n) at ({-\ra}, {-\dv});
        \coordinate[] (x0p) at ($({\ra}, {\dv}) + ({-\hsh}, {-\vsh})$);
        \coordinate[] (x0n) at ($({\ra}, {-\dv}) + ({-\hsh}, {-\vsh})$);
        \coordinate[] (x1p) at ({\dv}, {\rv});
        \coordinate[] (x1n) at ({-\dv}, {\rv});
        
        \coordinate[] (x2ip) at ({-\rt}, {\rv});
        \coordinate[] (x2in) at ({-\rt}, {-\rv});
        \coordinate[] (x0ip) at ($({\rt}, {\rv}) + ({-\hsh}, {-\vsh})$);
        \coordinate[] (x0in) at ($({\rt}, {-\rv}) + ({-\hsh}, {-\vsh})$);
        \coordinate[] (x1ip) at ({\rv}, 0);
        \coordinate[] (x1in) at ({-\rv}, 0);
        \coordinate[] (x1vn) at (0, {-\rv});
        \coordinate[] (x2i63p) at ($({-\rt}, 0)+\rv*({cos(\argAA)}, {sin(\argAA)})$);
        \coordinate[] (x2i63n) at ($({-\rt}, 0)+\rv*({cos(\argAB)}, {sin(\argAB)})$);
        \coordinate[] (x1i45p) at ($\rv*({cos(\argBA)}, {sin(\argBA)})$);
        \coordinate[] (x1i45n) at ($\rv*({cos(\argBB)}, {sin(\argBB)})$);
        \coordinate[] (x0i63p) at ($({\rt}, 0) + ({-\hsh}, {-\vsh})+\rv*({cos(\argAA)}, {sin(\argAA)})$);
        \coordinate[] (x0i63n) at ($({\rt}, 0) + ({-\hsh}, {-\vsh})+\rv*({cos(\argAB)}, {sin(\argAB)})$);
        \coordinate[] (x0i45p) at ($({\rt}, 0) + ({-\hsh}, {-\vsh})+\rv*({cos(\argBA)}, {sin(\argBA)})$);
        \coordinate[] (x0i45n) at ($({\rt}, 0) + ({-\hsh}, {-\vsh})+\rv*({cos(\argBB)}, {sin(\argBB)})$);
        
        \fill (x2) circle [radius=2pt, red];\draw ({-\rt}, 0.2) node[above]{$-1$};
        \fill (x1) circle [radius=2pt, red];\draw (0.3, 0.2) node[above]{$0$};
        \fill (x0) circle [radius=2pt, red];\draw (2, -1.3) node[above]{$1-i$};
        
        \draw[line width=0.4mm, gray, dashed] (ni) -- (x2);
        \draw[line width=0.4mm, gray, dashed] (x1) -- (0i);
        \draw[line width=0.4mm, gray, dashed] (x0) -- (pi);
        
        \draw[line width=0.3mm, red] ($(x2p)+(0,{\dd})$) to [out=85, in=180] ($(x2ip)+(0,{\dd})$);
        \draw[line width=0.3mm, red] ($(x2ip)+(0,{\dd})$) to [out=0, in=180] ($(x1vn)-(0,{\dd})$);
        \draw[line width=0.3mm, red] ($(x1vn)-(0,{\dd})$) to [out=0, in=270] ($(x1ip)-(0,{\dd})$);
        \draw[line width=0.3mm, red] ($(x1ip)-(0,{\dd})$) to [out=90, in=-5] ($(x1p)-(0,{\dd})$);
        
        \draw[line width=0.3mm, dashed, red] ($(x1n)-(0,{\dd})$) to [out=185, in=0] ($(x2in)+(0,{\dd})$);
        \draw[line width=0.3mm, dashed, red] ($(x2in)+(0,{\dd})$) to [out=180, in=275] ($(x2n)+(0,{\dd})$);

        \draw[line width=0.3mm, black] ($(x1p)+(0,{\dd})$) to [out=-5, in={-\argBB+10}] ($(x1i45p)+(0,{\dd})$);
        \draw[line width=0.3mm, black] ($(x1i45p)+(0,{\dd})$) to [out={-\argBA}, in={-\argBB+10}] ($(x0i45n)+(0,{\dd})$);
        \draw[line width=0.3mm, black] ($(x0i45n)+(0,{\dd})$) to [out={-\argBB+10}, in=180] ($(x0in)+(0,{\dd})$);
        \draw[line width=0.3mm, black] ($(x0in)+(0,{\dd})$) to [out=-10, in=265] (x0n);
        
        \draw[line width=0.3mm, dashed, black] (x0p) to [out=95, in={-\argBB}] (x0i45p);
        \draw[line width=0.3mm, dashed, black] (x0i45p) to [out={-\argBB}, in={-\argBA+5}] ($(x1i45n)+(0,{\dd})$);
        \draw[line width=0.3mm, dashed, black] ($(x1i45n)+(0,{\dd})$) to [out={-\argBB}, in=270] ($(x1in)+(0,{\dd})$);
        \draw[line width=0.3mm, dashed, black] ($(x1in)+(0,{\dd})$) to [out=90, in=185] ($(x1n)+(0,{\dd})$);

        \draw[line width=0.3mm, blue] ($(x2p)+({\dd}, 0)$) to [out=85, in=180] ($(x2ip)+({\dd}, 0)$);
        \draw[line width=0.3mm, blue] ($(x2ip)+({\dd}, 0)$) to [out=0, in={\argab}] ($(x2i63p)+({\dd}, 0)$);
        \draw[line width=0.3mm, blue] ($(x2i63p)+({\dd}, 0)$) to [out={-\argaa}, in={\argab+5}] ($(x0i63n)+({-\dd}, 0)$);
        \draw[line width=0.3mm, blue] ($(x0i63n)+({-\dd}, 0)$) to [out={-\argaa+5}, in=175] ($(x0in)+({-\dd}, 0)$);
        \draw[line width=0.3mm, blue] ($(x0in)+({-\dd}, 0)$) to [out=-5, in=265] ($(x0n)+({-\dd}, 0)$);
        
        \draw[line width=0.3mm, dashed, blue] ($(x0p)+({-\dd}, 0)$) to [out=95, in={-\argaa}] ($(x0i63p)+({-\dd}, 0)$);
        \draw[line width=0.3mm, dashed, blue] ($(x0i63p)+({-\dd}, 0)$) to [out=\argab, in={-\argaa}] ($(x2i63n)+({\dd}, 0)$);
        \draw[line width=0.3mm, dashed, blue] ($(x2i63n)+({\dd}, 0)$) to [out=\argab, in=275] ($(x2n)+({\dd}, 0)$);
        
        \end{tikzpicture}
        \subcaption{}
    \end{minipage}
    
    \caption{One-cycles on the WKB curves for the minimal chamber (a) and outside the minimal chamber (b).  
    Moving from the minimal chamber to the next chamber in the moduli space, a new cycle is involved in the BPS spectrum charge basis. In this figure, the dashed black lines starting from the zeros (black dots) indicate the branch cuts. The black and red cycles are $\hat{\gamma}_{1,1}$ and $\hat{\gamma}_{1,2}$ respectively. The blue cycle is $\hat{\gamma}_{1,1}+\hat{\gamma}_{1,2}$.} 
    \label{fig:cycle-wc}
\end{figure}
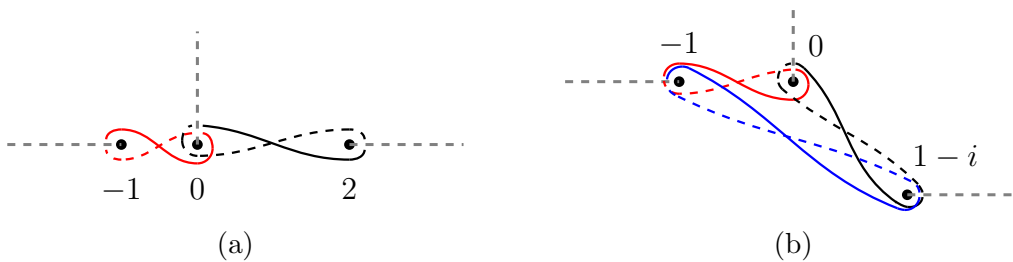
\begin{table}[H]
    \centering
    \small
    \begin{tabular}{r||r|r}
        $n$ & $\Pi_{\hat{\gamma}_{1,1}}^{(n)}$ & $(m_{1, 1}^{\Blue{{\rm n}}})^{(n - 1)}$ \\
        \hline
        $0$ & $1.874620612+0.3018501837i$ & $1.874620612+0.3018501837i$ \\
        $2$ & $0.1284063138+0.3047229198i$ & $0.1284063137+0.3047229201i$ \\
        $4$ & $0$ & $0$ \\
        $6$ & $5.769417182-3.128435216i$ & $5.769417207-3.128435225i$ \\
        $8$ & $174.0185754-30.03032475i$ & $174.0185763-30.03032474i$ \\
        $10$ & $0$ & $0$ \\
        $12$ & $-467246.4117-339241.7847i$ & $-467246.3493-339241.7425i$ \\
        $14$ & $-32138323.83-52653279.18i$ & $-32138271.57-52653196.97i$ \\
        \hline \hline
        $n$ & $e^{-\frac{\pi i}{3}(n-1)}\Pi_{\hat{\gamma}_{1,2}}^{(n)}$ & $(m_{1, 2}^{\rm n})^{(n - 1)}$ \\
        \hline
        $0$ & $0.2197744901+1.094413217i$ & $0.2197744901+1.094413217i$ \\
        $2$ & $-0.1284063138-0.3047229198i$ & $-0.1284063137-0.3047229201i$ \\
        $4$ & $0$ & $0$ \\
        $6$ & $-5.769417182+3.128435216i$ & $-5.769417207+3.128435225i$ \\
        $8$ & $-174.0185754+30.03032475i$ & $-174.0185763+30.03032474i$ \\
        $10$ & $0$ & $0$ \\
        $12$ & $467246.4117+339241.7847i$ & $467246.3493+339241.7425i$ \\
        $14$ & $32138323.83+52653279.18i$ & $32138271.57+52653196.97i$ \\
    \end{tabular}
    \caption{$\Pi_{\hat{\gamma}_{1,k}}^{(n)}$ and $(m_{1,k}^{\rm n})^{(n-1)}$ ($k=1,2$) for the third order ODE with $p(x) = -x^3-ix^2+(1-i)x$. The branch points are $x_0 = 1 - i$, $x_1 = 0$, $x_2 = -1$. See Fig. \ref{fig:cycle-wc} (b).}
    \label{tab:A2A2_next}
\end{table}
\begin{table}[h]
    \centering
    \small
    \begin{tabular}{r||r|r}
        $n$ & $\Pi_{\hat{\gamma}_{1,12}}^{(n)}$ & $(m_{12}^{{\rm n}})^{(n - 1)}$ \\
        \hline
        $0$ & $2.932297506 + 0.6587265011i$ & $2.932297506 + 0.6587265011i$ \\
        $2$ & $0.3281009466 + 0.04115833014i$ & $0.3281009468 + 0.04115833032i$ \\
        $4$ & $0$ & $0$ \\
        $6$ & $5.594012963 + 3.432244237i$ & $5.594012983 + 3.432244253i$ \\
        $8$ & $61.00226359 - 165.7196694i$ & $61.00226404 - 165.7196701i$ \\
        $10$ & $0$ & $0$ \\
        $12$ & $60168.79778 - 574268.1548i$ & $60168.79043 - 574268.0568i$ \\
        $14$ & $-61668239.29 + 1505965.287i$ & $-61668121.03 + 1505960.255i$
    \end{tabular}
    \caption{$\Pi_{\hat{\gamma}_{1,12}}^{(n)}$ and $(m_{12}^{\rm n})^{(n-1)}$ for the third order ODE with $p(x) = -x^3-ix^2+(1-i)x$.}
    \label{tab:A2A2_next_2}
\end{table}
The TBA equations (\ref{eq:new-tba}) thus describe the WKB periods in the next chamber outside the minimal one.
We can also perform further wall-crossing of the TBA equations to go to the maximal chamber, which includes the case where the potential is monomial \cite{ItKoKuSh3}.

\section{Conclusions and Discussion}
\label{sec:conclusions}
In this paper, we studied the WKB periods of the ODE which is a higher order generalization of the Schr\"odinger equation.
The quantum corrections of the WKB periods were obtained by solving the Riccati equation recursively.
These corrections were expressed by applying the Picard-Fuchs operators to the classical SW periods.
We also defined the Y-functions from the Wronskians and derived the Y-system and the TBA equations.
Based on the results of the Schr\"odinger type ODE and the third order ODE, we proposed a formula which relates the WKB periods to the Y-functions for the higher order ODE with the quadratic potential.
We checked the formula numerically.
We also derived a non-trivial relation between the classical WKB period and the second order quantum correction.
The relation takes the form of the PNP relation and has been re-derived from the WKB side.
It thus provides the other non-trivial test of the identification between the Y-functions and the WKB periods.

In the exact WKB analysis, one needs to perform the Borel resummation of the WKB periods.
The Y-function should correspond to the Borel resummation of the WKB period.
In our TBA equations, the poles of the kernel function have a rich structure.
The identification implies that the WKB periods studied in our paper are Borel summable for positive real $\epsilon$, but are Borel non-summable at $\arg(\epsilon)=\pm \pi/h$ along which the poles in the kernel function are located.
It would be interesting to make an extensive study in this direction.

It is also interesting to consider a generalization to the higher order polynomial potential.
In this case, since the phase of the mass parameters are not necessarily the same and the integration path of the TBA equations could cross the pole, more complicated wall-crossing phenomena may occur \cite{Ito:2018eon}. 
We have studied the third order ODE with the cubic potential in detail. We have seen an agreement between the WKB periods and the Y-functions before/after the wall-crossing.

There is another direction to generalize the present work.
The ODE studied in this paper is obtained from the $A_r^{(1)}$-type linear problem without monodromies around the origin, which is associated with the affine Toda equation.
The WKB analyses for the linear problem associated with the affine Toda equation for other affine Lie algebras and/or with monodromies are also interesting in the viewpoint of the analytic structure of the Y-functions.

The ODE studied in this paper is regarded as the quantum Seiberg-Witten curve of the $(A_r, A_1)$-type Argyres-Douglas theory which is dual to the $(A_1, A_r)$-type Argyres-Douglas theory.
Our TBA equations have the same form as the ones of the $(A_1, A_r)$-type ODE, which has been studied in \cite{Dorey:2000ma, Ito:2017ypt}.
This can be interpreted as the duality between the $(A_r, A_1)$-type AD theory and the $(A_1, A_r)$-type AD theory in the NS limit of the Omega background.
It is interesting to study the quantum SW curves for $(D_r, A_1)$ and $(E_r, A_1)$ types by using the WKB analysis and the TBA equations (see \cite{Longhi:2016rjt} for a related work).

\subsection*{Acknowledgements}
We would like to thank Davide Fioravanti, Daniele Gregori,  Hao Ouyang and Marco Rossi for useful discussions.
The work of K.I. is supported in part by Grant-in-Aid for Scientific Research 21K03570, 18K03643 and 17H06463 from Japan Society for the Promotion of Science (JSPS).
The work of H.S. is supported by the grant ``Exact Results in Gauge and String Theories'' from the Knut and Alice Wallenberg foundation.

\appendix
\section{Higher order terms in the WKB expansion}
\label{sec:wkb}
In this appendix, we present the coefficients $S_{2i}$ $(i=1,2,3,4)$ in the WKB expansions \eqref{eq:formal_series_exp_S} of $S$ for the $(r+1)$-th order ODE.
$S_{2i}$ are calculated up to total derivatives:
\begin{equation}
    \begin{aligned}
        S_2&=\frac{h^2-1}{48}\frac{\partial^2 S_0}{S_0^2},\\
        S_4&=\frac{(h^2-9)(h^2-1)}{768}\frac{(\partial^2S_0)^2}{S_0^5}-\frac{(h^2-9)(h^2-1)}{5120} \frac{\partial^4 S_0}{S_0^4},\\
        S_6&=a_6\frac{(\partial^2 S_0)^3}{S_0^8}+b_6 \frac{(\partial^3 S_0)^2}{S_0^7}+c_6 \frac{ \partial^2S_0 \partial^4S_0}{S_0^7}+d_6 \frac{\partial^6S_0}{S_0^6},\\
        S_8&=a_8 \frac{(\partial^2 S_0)^4}{S_0^{11}}+b_8 \frac{\partial^2 S_0 (\partial^3 S_0)^2}{S_0^{10}}+c_8 \frac{(\partial^2 S_0)^2 \partial^4 S_0}{S_0^{10}}\\
        &+d_8\frac{(\partial^4 S_0)^2}{S_0^{9}}+e_8  \frac{\partial^3 S_0 \partial^5 S_0}{S_0^{9}}+f_8 \frac{\partial^2 S_0 \partial^6 S_0}{S_0^{9}}+g_8\frac{\partial^8 S_0}{S_0^8},
    \end{aligned}
\end{equation}
where
\begin{equation}
    \begin{aligned}
        a_6&=\frac{(h^2-5^2)(h^2-1) (401 h^2-2081)}{1327104},\\
        b_6&=-\frac{(h^2-5^2)(h^2-1)^2}{387072},\\
        c_6&=-\frac{(h^2-5^2)(h^2-1)(11h^2-59)}{147456},\\
        d_6&=\frac{(h^2-5^2)(h^2-1)(11h^2-59)}{6193152},
    \end{aligned}
\end{equation}
and
\begin{equation}
    \begin{aligned}
        a_8&=\frac{(h^2-7^2)(h^2-1) (23490253-7310426 h^2+452173 h^4)}{3503554560},\\
        b_8&=-\frac{(h^2-7^2)(h^2-1)^2 (-3649+409h^2)}{364953600},\\
        c_8&=-\frac{(h^2-7^2)(h^2-1) (4674697-1436114 h^2+87817 h^4)}{1946419200},\\
        d_8&=\frac{(h^2-7^2)(h^2-1)(2792063-847678 h^2+51455 h^4)}{35035545600},\\
        e_8&=\frac{(h^2-7^2)(h^2-1)^2(-509+53 h^2)}{547430400},\\
        f_8&=\frac{(h^2-7^2)(h^2-1) (73943-22510 h^2+1367 h^4)}{973209 600},\\
        g_8&=-\frac{(h^2-7^2)(h^2-1) (73943-22510 h^2+1367 h^4)}{93428121600}.
    \end{aligned}
\end{equation}

\section{The PF operators and quantum corrections}
\label{sec:PF_op_q_corrections}
In the following, we show the Picard-Fuchs operators and the ratio of the quantum corrections to the classical SW periods up to $\epsilon^{16}$ for the ODE of order from four to seven.
For $r=2k+1$ $(k=0,1,\dots)$, $l_n$ becomes
\begin{equation}
    \{l_{2n}\}_{n=1,2,\dots}=\{2k+1, 2k-1,\dots,1,2k+1,\dots\}.
\end{equation}
For $r=2k$ $(k=0,1,\dots)$, $l_n$ becomes
\begin{equation}
    \{l_{2n}\}_{n=1,2,\dots}=\{2k, 2k-2,\dots,0,2k-1,2k-3,\dots,1,2k,\dots\}.
\end{equation}

\subsection{The fourth order ODE}
The PF operators are
\begin{equation}
    \begin{aligned}
         \mathcal{O}^{\mathrm{PF}}_2&=\frac{u_0}{3}\partial_{u_2}^2, & 
         \mathcal{O}^{\mathrm{PF}}_4&=-\frac{11 u_0^2}{120}\partial_{u_2}^3, & 
         \mathcal{O}^{\mathrm{PF}}_6&=\frac{61 u_0^3}{1080}\partial_{u_2}^5, \\
         \mathcal{O}^{\mathrm{PF}}_8&=-\frac{353 u_0^4}{8064}\partial_{u_2}^6, &
         \mathcal{O}^{\mathrm{PF}}_{10}&=\frac{11099 u_0^5}{362880}\partial_{u_2}^8, &
         \mathcal{O}^{\mathrm{PF}}_{12}&=-\frac{49707277 u_0^6}{2075673600}\partial_{u_2}^9, \\
         \mathcal{O}^{\mathrm{PF}}_{14}&=\frac{4828591 u_0^7}{230630400}\partial_{u_2}^{11}, &
         \mathcal{O}^{\mathrm{PF}}_{16}&=-\frac{4477909193 u_0^8}{219967488000}\partial_{u_2}^{12}.
    \end{aligned}
\end{equation}
The ratio of the quantum corrections to the classical SW periods are
\begin{equation}
    \begin{aligned}
        &\frac{\Pi^{(2)}_{\gamma}}{\hat{\Pi}^{(0),3}_{\gamma}}=\frac{5 u_0}{48 u_2^2}, & 
        &\frac{\Pi^{(4)}_{\gamma}}{\hat{\Pi}^{(0),1}_{\gamma}}=-\frac{11 u_0^2}{512 u_2^3},\\
        &\frac{\Pi^{(6)}_{\gamma}}{\hat{\Pi}^{(0),3}_{\gamma}}=-\frac{4697 u_0^3}{73728 u_2^5}, & 
        &\frac{\Pi^{(8)}_{\gamma}}{\hat{\Pi}^{(0),1}_{\gamma}}=\frac{1170195 u_0^4}{3670016 u_2^6},\\
        &\frac{\Pi^{(10)}_{\gamma}}{\hat{\Pi}^{(0),3}_{\gamma}}=\frac{266764465 u_0^5}{75497472 u_2^8}, & 
        &\frac{\Pi^{(12)}_{\gamma}}{\hat{\Pi}^{(0),1}_{\gamma}}=-\frac{122528437805 u_0^6}{2952790016 u_2^9},\\
        &\frac{\Pi^{(14)}_{\gamma}}{\hat{\Pi}^{(0),3}_{\gamma}}=-\frac{61815211551765 u_0^7}{55834574848 u_2^{11}}, & 
        &\frac{\Pi^{(16)}_{\gamma}}{\hat{\Pi}^{(0),1}_{\gamma}}=\frac{15168742752828973 u_0^8}{549755813888 u_2^{12}}.
    \end{aligned}
\end{equation}

\subsection{The fifth order ODE}
The PF operators are
\begin{equation}
    \begin{aligned}
        \mathcal{O}^{\mathrm{PF}}_2&=\frac{5 u_0}{12}\partial_{u_2}^2, & 
        \mathcal{O}^{\mathrm{PF}}_4&=-\frac{13 u_0^2}{72}\partial_{u_2}^3, &
        \mathcal{O}^{\mathrm{PF}}_8&=-\frac{3889 u_0^4}{18144}\partial_{u_2}^6, \\
        \mathcal{O}^{\mathrm{PF}}_{10}&=\frac{8177 u_0^5}{28512}\partial_{u_2}^7, &
        \mathcal{O}^{\mathrm{PF}}_{12}&=-\frac{5801857 u_0^6}{13837824}\partial_{u_2}^9, &
        \mathcal{O}^{\mathrm{PF}}_{14}&=\frac{6172661 u_0^7}{8491392}\partial_{u_2}^{10}.
    \end{aligned}
\end{equation}
The ratio of the quantum corrections to the classical SW periods are
\begin{equation}
    \begin{aligned}
        &\frac{\Pi^{(2)}_{\gamma}}{\hat{\Pi}^{(0),4}_{\gamma}}= \frac{13 u_0}{80 u_2^2}, & 
        &\frac{\Pi^{(4)}_{\gamma}}{\hat{\Pi}^{(0),2}_{\gamma}}=-\frac{143 u_0^2}{8000 u_2^3},\\
        &\frac{\Pi^{(8)}_{\gamma}}{\hat{\Pi}^{(0),3}_{\gamma}}=-\frac{306425977 u_0^4}{672000000 u_2^6}, & 
        &\frac{\Pi^{(10)}_{\gamma}}{\hat{\Pi}^{(0),1}_{\gamma}}=\frac{39003856619 u_0^5}{2880000000 u_2^7},\\
        &\frac{\Pi^{(12)}_{\gamma}}{\hat{\Pi}^{(0),4}_{\gamma}}=\frac{1965108059811387 u_0^6}{5632000000000 u_2^9}, & &\frac{\Pi^{(14)}_{\gamma}}{\hat{\Pi}^{(0),2}_{\gamma}}=-\frac{57186073300864563 u_0^7}{16640000000000 u_2^{10}}.
    \end{aligned}
\end{equation}

\subsection{The sixth order ODE}
The PF operators are
\begin{equation}
    \begin{aligned}
        \mathcal{O}^{\mathrm{PF}}_2&=\frac{u_0}{2}\partial_{u_2}^2, & 
        \mathcal{O}^{\mathrm{PF}}_6&=\frac{5135 u_0^3}{12096}\partial_{u_2}^4, &
        \mathcal{O}^{\mathrm{PF}}_8&=-\frac{75709 u_0^4}{96768}\partial_{u_2}^6, \\
        \mathcal{O}^{\mathrm{PF}}_{12}&=-\frac{50444678155 u_0^6}{11955879936}\partial_{u_2}^8, &
        \mathcal{O}^{\mathrm{PF}}_{14}&=\frac{271675168801 u_0^7}{23911759872}\partial_{u_2}^{10}.
    \end{aligned}
\end{equation}
The ratio of the quantum corrections to the classical SW periods are
\begin{equation}
    \begin{aligned}
        &\frac{\Pi^{(2)}_{\gamma}}{\hat{\Pi}^{(0),5}_{\gamma}}=\frac{2 u_0}{9 u_2^2}, & 
        &\frac{\Pi^{(6)}_{\gamma}}{\hat{\Pi}^{(0),1}_{\gamma}}=-\frac{5135 u_0^3}{17496 u_2^4},\\
        &\frac{\Pi^{(8)}_{\gamma}}{\hat{\Pi}^{(0),5}_{\gamma}}=-\frac{4163995 u_0^4}{1102248 u_2^6}, & 
        &\frac{\Pi^{(12)}_{\gamma}}{\hat{\Pi}^{(0),1}_{\gamma}}=\frac{4792244424725 u_0^6}{3367210176 u_2^8},\\
        &\frac{\Pi^{(14)}_{\gamma}}{\hat{\Pi}^{(0),5}_{\gamma}}=\frac{2655624775029775 u_0^7}{35814871872 u_2^{10}}.
    \end{aligned}
\end{equation}
The corrections for order $n=6k+4$ $(k=0,1,\dots)$ vanish for the quadratic potential, but do not vanish for the general potential.

\subsection{The seventh order ODE}
The PF operators are
\begin{equation}
    \begin{aligned}
        \mathcal{O}^{\mathrm{PF}}_2&=\frac{7 u_0}{12}\partial_{u_2}^2, & 
        \mathcal{O}^{\mathrm{PF}}_4&=-\frac{119 u_0^2}{240}\partial_{u_2}^3, \\
        \mathcal{O}^{\mathrm{PF}}_6&=\frac{29 u_0^3}{30}\partial_{u_2}^4, &
        \mathcal{O}^{\mathrm{PF}}_{10}&=\frac{765289 u_0^5}{95040}\partial_{u_2}^7, \\
        \mathcal{O}^{\mathrm{PF}}_{12}&=-\frac{171906011 u_0^6}{5702400}\partial_{u_2}^8, &
        \mathcal{O}^{\mathrm{PF}}_{14}&=\frac{434398427 u_0^7}{3706560}\partial_{u_2}^9, \\
        \mathcal{O}^{\mathrm{PF}}_{16}&=-\frac{3131950604279 u_0^8}{5930496000}\partial_{u_2}^{11}.
    \end{aligned}
\end{equation}
The ratio of the quantum corrections to the classical SW periods are
\begin{equation}
    \begin{aligned}
        &\frac{\Pi^{(2)}_{\gamma}}{\hat{\Pi}^{(0),6}_{\gamma}}=\frac{95 u_0}{336 u_2^2}, & 
        &\frac{\Pi^{(4)}_{\gamma}}{\hat{\Pi}^{(0),4}_{\gamma}}=\frac{221 u_0^2}{6272 u_2^3},\\ 
        &\frac{\Pi^{(6)}_{\gamma}}{\hat{\Pi}^{(0),2}_{\gamma}}=-\frac{168113u_0^3}{384160 u_2^4}, &
        &\frac{\Pi^{(10)}_{\gamma}}{\hat{\Pi}^{(0),5}_{\gamma}}=-\frac{428983295585 u_0^5}{2891341824 u_2^7},\\
        &\frac{\Pi^{(12)}_{\gamma}}{\hat{\Pi}^{(0),3}_{\gamma}}=\frac{63908958440313493 u_0^6}{37402397835264 u_2^8}, & 
        &\frac{\Pi^{(14)}_{\gamma}}{\hat{\Pi}^{(0),1}_{\gamma}}=\frac{5422964705164303925 u_0^7}{17189990924288 u_2^9},\\
        &\frac{\Pi^{(16)}_{\gamma}}{\hat{\Pi}^{(0),6}_{\gamma}}=\frac{50642618652822320762759 u_0^8}{1692552952545280 u_2^{11}}.
    \end{aligned}
\end{equation}

\section{Expansion of T-operator}\label{sec:T-expansion}
In \cite{Bazhanov:2001xm}, the authors studied the integrable structure of the $W_3$ conformal field theory with the central charge $c=50-24(g+g^{-1})$, and constructed the ${\bf T}$ and ${\bf Q}$ operators acting on the modules ${\cal V}_{\Delta_2, \Delta_3}$ with the highest weights
\begin{equation}
        \Delta_{2}=\frac{p_{1}^{2}+p_{2}^{2}}{g}+\frac{c-2}{24},\quad\Delta_{3}=\frac{2p_{2}(p_{2}^{2}-3p_{1}^{2})}{(3g)^{3/2}}.
\end{equation}
At large spectral parameter $t$, the $\mathbf{T}$-operator is expanded in terms of the integrals of motions ${\bf I}_{k}$:
\begin{equation}
    \begin{aligned}
        \log\mathbf{T}(t)&\sim mt^{\frac{1}{3(1-g)}}{\bf I}-2\sum_{n=1}^{\infty}C_{2n}\cos(\frac{\pi n}{3})t^{-\frac{2n}{3(1-g)}}{\bf I}_{2n}\\
        &\quad+2i\sum_{n=1}^{\infty}C_{2n-1}\sin(\frac{\pi(2n-1)}{6})t^{-\frac{2n-1}{3(1-g)}}{\bf I}_{2n-1},
    \end{aligned}
\end{equation}
where $m=\frac{2\pi\Gamma(\frac{2}{3}-\frac{\xi}{3})}{\Gamma(1-\frac{\xi}{3})\Gamma(\frac{2}{3})}\Big(\Gamma(1-g)\Big)^{\frac{1}{1-g}}$ and $\xi=\frac{g}{1-g}$. The details of the coefficients $C_k$ can be found in \cite{Bazhanov:2001xm}. The first non-zero vacuum eigenvalues of the integrals of motions ${\bf I}_{k}$ are
\begin{equation}
    \begin{aligned}
        I_{1}^{({\rm vac})}&=\Delta_{2}-\frac{c}{24},\\I_{2}^{({\rm vac})}&=\Delta_{3},\\I_{3}^{({\rm vac})}&=\Delta_{3}(\Delta_{2}-\frac{c+6}{24}),\\I_{4}^{({\rm vac})}&=\Delta_{2}^{3}+\frac{4\Delta_{3}^{2}}{3}-\frac{c+8}{8}\Delta_{2}^{2}+\frac{(c+2)(c+15)}{192}\Delta_{2}-\frac{c(c+23)(7c+30)}{96768}.
    \end{aligned}
\end{equation}
Moreover, the ODE \eqref{eq:Ar_ODE} with $r=2$ is proposed to correspond to the vacuum state of $W_3$ integrable model with $g=3/5$, $\Delta_2=-1/5$ and $\Delta_3=0$ \cite{Bazhanov:2001xm}.
To test this correspondence, let us compare the expansion of $T$-operator on the vacuum state with the expansion of the WKB periods $\Pi_{\gamma_1}$ at large $\theta$:
\begin{equation}
    \Pi_{\gamma_1} = \Pi_{\gamma_1}^{(0)}e^{\theta}+\Pi_{\gamma_1}^{(2)}e^{-\theta}+\Pi_{\gamma_1}^{(6)}e^{-5\theta}+\cdots,
\end{equation}
which corresponds with $\log Y_{1,1}$ and thus is expected to related with $\log {\bf T}$ on the vacuum state.
Keeping this in mind, it is easy to find
\begin{equation}
    \log T(t) = \Pi_{\gamma_1},
\end{equation}
under the identification
\begin{equation}
    t=\frac{(-1)^{\frac{4}{5}}u_2}{5^{\frac{6}{5}}\Gamma\qty(\frac{2}{5})^3u_0^{\frac{3}{5}}}e^{\frac{6\theta}{5}}.    
\end{equation}

\end{document}